\documentclass[showpacs,amsfonts,aps,prc,floatfix,eqsecnum,groupedaddress,]%
{revtex4}
\usepackage{amsmath,amssymb,amsthm}
\usepackage{graphicx}
\usepackage{bm}

\DeclareMathAlphabet{\mathpzc}{OT1}{pzc}{m}{it}

\begin{document}

\renewcommand{\textfraction}{0.00}

% Useful macros:

\newcommand{\vAi}{{\cal A}_{i_1\cdots i_n}} 
\newcommand{\vAim}{{\cal A}_{i_1\cdots i_{n-1}}} 
\newcommand{\vAbi}{\bar{\cal A}^{i_1\cdots i_n}}
\newcommand{\vAbim}{\bar{\cal A}^{i_1\cdots i_{n-1}}}
\newcommand{\htS}{\hat{S}} 
\newcommand{\htR}{\hat{R}}
\newcommand{\htB}{\hat{B}} 
\newcommand{\htD}{\hat{D}}
\newcommand{\htV}{\hat{V}} 
\newcommand{\cT}{{\cal T}} 
\newcommand{\cM}{{\cal M}} 
\newcommand{\cMs}{{\cal M}^*}
\newcommand{\vk}{\vec{\mathbf{k}}}
\newcommand{\bk}{\bm{k}}
\newcommand{\kt}{\bm{k}_\perp}
\newcommand{\kp}{k_\perp}
\newcommand{\km}{k_\mathrm{max}}
\newcommand{\vl}{\vec{\mathbf{l}}}
\newcommand{\bl}{\bm{l}}
\newcommand{\bK}{\bm{K}} 
\newcommand{\bb}{\bm{b}} 
\newcommand{\qm}{q_\mathrm{max}}
\newcommand{\vp}{\vec{\mathbf{p}}}
\newcommand{\bp}{\bm{p}} 
\newcommand{\vq}{\vec{\mathbf{q}}}
\newcommand{\bq}{\bm{q}} 
\newcommand{\qt}{\bm{q}_\perp}
\newcommand{\qp}{q_\perp}
\newcommand{\bQ}{\bm{Q}}
\newcommand{\vx}{\vec{\mathbf{x}}}
\newcommand{\bx}{\bm{x}}
\newcommand{\tr}{{{\rm Tr\,}}} 

\newcommand{\beq}{\begin{equation}}
\newcommand{\eeq}[1]{\label{#1} \end{equation}} 
\newcommand{\ee}{\end{equation}}
\newcommand{\bea}{\begin{eqnarray}} 
\newcommand{\eea}{\end{eqnarray}}
\newcommand{\beqar}{\begin{eqnarray}} 
\newcommand{\eeqar}[1]{\label{#1}\end{eqnarray}}
 
\newcommand{\half}{{\textstyle\frac{1}{2}}} 
\newcommand{\ben}{\begin{enumerate}} 
\newcommand{\een}{\end{enumerate}}
\newcommand{\bit}{\begin{itemize}} 
\newcommand{\eit}{\end{itemize}}
\newcommand{\bc}{\begin{center}} 
\newcommand{\ec}{\end{center}}
\newcommand{\bra}[1]{\langle {#1}|}
\newcommand{\ket}[1]{|{#1}\rangle}
\newcommand{\norm}[2]{\langle{#1}|{#2}\rangle}
\newcommand{\brac}[3]{\langle{#1}|{#2}|{#3}\rangle} 
\newcommand{\hilb}{{\cal H}} 
\newcommand{\pleft}{\stackrel{\leftarrow}{\partial}}
\newcommand{\pright}{\stackrel{\rightarrow}{\partial}}

%%%%%%%%%%%%%%%%%%%%%%%%% Front Matter %%%%%%%%%%%%%%%%%%%%%%%%%%%%%%%
\title{Radiative heavy quark energy loss in a dynamical QCD medium}
\author{Magdalena Djordjevic}
%\email[Correspond to\ ]{magda@mps.ohio-state.edu}
\affiliation{Physics Department, The Ohio State University,
%  191 W. Woodruff Avenue,
Columbus, OH 43210, USA}
\author{Ulrich Heinz}
\affiliation{Physics Department, The Ohio State University,
%  191 W. Woodruff Avenue,
Columbus, OH 43210, USA}

\begin{abstract}
The computation of radiative energy loss in a dynamically screened QCD 
medium is a key ingredient for obtaining reliable predictions for jet 
quenching in ultra-relativistic heavy ion collisions. We calculate, to 
first order in the opacity, the energy loss suffered by a heavy quark 
traveling through an infinite and time-independent QCD medium and show 
that the result for a dynamical medium is almost twice that obtained
previously for a medium consisting of randomly distributed static 
scattering centers. A quantitative description of jet suppression in RHIC 
and LHC experiments thus must correctly account for the dynamics of the 
medium's constituents. 
\end{abstract}

\date{\today} 
 
\pacs{25.75.-q, 25.75.Nq, 12.38.Mh, 12.38.Qk} 

\maketitle
%%%%%%%%%%%%%%%%%%%%%%%%%%%%%%%%%%%%%%%%%%%%%%%%%%%%%%%%%%%%%%%%%%%%%%%

%%%%%%%%%%%%%%%%%%%%%%%%%%%%%%%%%%%%%%%%%%%%%%%%%%%%%%%%%%%%%%%%%%%%%%%
\section{Introduction}
\label{sec1}
%%%%%%%%%%%%%%%%%%%%%%%%%%%%%%%%%%%%%%%%%%%%%%%%%%%%%%%%%%%%%%%%%%%%%%%

Studying the suppression pattern of high transverse momentum hadrons 
is a powerful tool to map out the density of a QCD plasma created in 
ultra-relativistic heavy ion collisions 
\cite{Gyulassy_2002,Gyulassy:1990bh,Gyulassy:1991xb}. This suppression 
(called jet quenching) results from the energy loss of high energy 
partons moving through the plasma \cite{MVWZ:2004,BDMS,BSZ,KW:2004}. Recent 
non-photonic single electron data~\cite{Adler:2005xv,elecQM05_STAR} (which 
present an indirect probe of heavy quark energy loss) showed that radiative 
energy loss alone can not explain the results as long as realistic 
parameter values are used \cite{Djordjevic:2005db}. Inclusion of 
collisional energy loss \cite{Mustafa,Dutt-Mazumder,MD_Coll,Adil}
improves the agreement with available data \cite{WHDG}, but still does
not yield a perfect description.

The currently available studies suffer from one crucial drawback: The medium 
induced radiative energy loss is computed in a QCD medium consisting
of randomly distributed but static scattering centers (``static QCD 
medium''). In such a medium the collisional energy loss is exactly 
zero. This approximation was motivated by early estimates 
\cite{Bjorken:1982tu,TG,BT,Wang:1994fx,Mustafa:1997pm,Lin:1997cn}, 
which indicated that the typical collisional energy loss should be
small compared to the radiative one. However, recent calculations
\cite{Mustafa,Dutt-Mazumder,MD_Coll,Adil} showed that the collisional 
contribution is important and comparable to the radiative energy loss.
The static approximation is thus qualitatively wrong as far as the 
computation of collisional energy loss is concerned and should therefore
also be revisited in the context of radiative energy loss.

In this paper, we report on a first important step, the calculation of
heavy quark radiative energy loss in an infinite and time-independent 
QCD medium consisting of dynamical constituents. By comparing with the
static medium calculation, this permits us to qualitatively assess the 
importance of dynamical effects on radiative energy loss. The more 
demanding problem of including finite medium size corrections, {\it i.e.} 
the Landau-Pomeranchuk-Migdal (LPM) effect, will be left to a future
study.

Here is the outline of our paper: In Section~\ref{sec2} we compute, to first 
order in the opacity, the radiative energy loss in an infinite, dynamical 
QCD medium. In Section~\ref{sec3} we obtain the corresponding result in the 
static approximation. While the analytical results in both cases lead to 
formally very similar expressions, they give remarkably different numerical 
values for the energy loss. These are presented in Section~\ref{sec4}. We 
will see that a dynamical medium leads to approximately twice the radiative 
energy loss obtained in the static approximation. In Section~\ref{sec5} we 
present a short summary and conclude that representing the QCD medium by a 
random ensemble of static scattering centers is not a good approximation for 
RHIC and LHC phenomenology. Some technical steps of our calculation are 
reproduced in the Appendix. We will use the following notation for
4-vectors: $k=(k_0,\vk)=(k_0,k_z,\bk)$, i.e. $\vk$ (with an explicit 
vector superscript) describes a 3-vector while $\bk$ (without a vector 
superscript) denotes the 2-vector transverse to the direction of motion 
$z$ of the heavy quark. Correspondingly $d^3k\equiv dk_z\,d^2k$.

%%%%%%%%%%%%%%%%%%%%%%%%%%%%%%%%%%%%%%%%%%%%%%%%%%%%%%%%%%%%%%%%%%%%%%%%
\section{Radiative energy loss in a dynamical QCD medium} 
\label{sec2}
%%%%%%%%%%%%%%%%%%%%%%%%%%%%%%%%%%%%%%%%%%%%%%%%%%%%%%%%%%%%%%%%%%%%%%%%

In this Section we compute the medium induced radiative energy loss 
for a heavy quark to first order in the opacity. For simplicity we 
consider an infinite QCD medium and assume that the on-shell heavy 
quark is produced at time $x_0=-\infty$. In this medium we compute the 
radiative energy loss per unit length, $\frac{dE}{dL}$. For phenomenological
applications in heavy-ion collisions one would, as a first approximation, 
use this result and simply multiply it with the effective thickness $L$ 
of the medium to calculate the total energy loss. A more rigorous 
derivation would have to start with a finite size medium from the 
beginning; we leave this for the future.

Medium induced radiative energy loss is caused by the radiation of one 
or more gluons induced by collisional interactions between the quark of 
interest and partons in the medium. The energy loss rate can be expanded
in the number of scattering events suffered by the heavy quark which is
equivalent to an expansion in powers of the opacity. For a finite medium,
the opacity is given by the product of the density of the medium with the 
scattering cross section, integrated along the path of the heavy quark.
The lowest (first) order contribution corresponds to one collisional
interaction with the medium, accompanied by emission of a single gluon. 
We adopt this as a definition of the ``first order in opacity'' also for 
the infinite medium. 

For a medium consisting of dynamical quarks and gluons in thermal 
equilibrium, the corresponding energy loss contribution involves
two cut Hard-Thermal Loop (HTL) gluon propagators. The associated
Feynman diagrams are plotted in Figs.~\ref{DiagM10}-\ref{DiagM12} 
and computed in Appendices \ref{appb}-\ref{appd}. The diagrams represent 
an on-shell heavy quark with momentum $p'$ which (in arbitrary order) 
exchanges a virtual gluon of momentum $q$ with a parton in the medium 
and radiates a gluon with momentum $k$. The heavy quark emerges with 
(measured) momentum $p$. Since the exchanged gluon momentum is space-like 
\cite{MD_Coll,BT,BT_fermions}), only the Landau damping contribution 
($q_0 \le |\vq|$) to the cut HTL effective gluon propagator $D(q)$
needs to be taken into account \cite{MD_Coll,TG,BT}. 

The radiated gluon has timelike momentum $k=(\omega,\vk)$, so only the 
quasi-particle contribution at $\omega> |\vk|$ from the cut gluon propagator 
$D(k)$ contributes \cite{DG_TM,Kapusta,Le_Bellac}. Energy and momentum 
conservation requires $p'=p+k+q$. Since our focus is on heavy quarks with 
mass $M\gg gT$, we neglect the thermal shift of the heavy 
quark mass.

The effective gluon propagator has both transverse and longitudinal 
contributions
\cite{Kalash,Klimov,Weldon,Heinz_AP,Pisarski:cs,Rebhan,Gyulassy_Selikhov}. 
The 1-HTL gluon propagator has the form
\beq
 i D^{\mu\nu}(l)=
 \frac{P^{\mu \nu }(l)}{l^2{-}\Pi_T (l)} + 
 \frac{Q^{\mu \nu }(l)}{l^2{-}\Pi_L (l)}\,,
\eeq{dmnMed}
where $l=(l_0, \vl)$ is the 4-momentum of the gluon and 
$P_{\mu \nu}(l)$ and $Q_{\mu \nu}(l)$ are the transverse and
longitudinal projectors, respectively. The transverse and longitudinal 
HTL gluon self energies $\Pi_T$ and $\Pi_L$ are given 
by \cite{Heinz_AP}
\bea
\label{PiT}
\Pi_T (l) &=& \mu^2 \left[ \frac{y^2}{2} + \frac{y (1{-}y^2)}{4} 
\ln\left(\frac{y{+}1}{y{-}1}\right)\right],
\qquad
%\\
\Pi_L (l) = \mu^2 \left[ 1 - y^2 - \frac{y (1{-}y^2)}{2} 
\ln\left(\frac{y{+}1}{y{-}1}\right)\right],
%\label{PiL}
\eea
where $y \equiv l_0/|\vl|$ and $\mu=gT\sqrt{N_c/3+N_f/6}$ is the
Debye screening mass.

While the results obtained in this paper are gauge invariant 
\cite{BT_fermions}, we present the calculation for simplicity in 
Coulomb gauge. In this gauge the only nonzero terms in the transverse 
and longitudinal projectors are
\bea
P^{i j} (l) &=& -\delta^{ij} + \frac{l^i  l^j}{\vl^2},
\label{PQmunu}
%\\
\qquad\qquad
Q^{00}(l) = -\frac{l^2}{\vl^2} = 1 -\frac{l_0^2}{\vl^2} = 1-y^2.
%\label{Q00}
\eea

As in~\cite{Gyulassy_Wang,GLV,Wiedemann,WW,DG_Ind,ASW,MD_TR}, we assume 
validity of the soft gluon and soft rescattering approximations (see 
Appendix~\ref{appa} for details). With these assumptions we compute
in Appendices \ref{appb}-\ref{appd} the diagrams $M_{1,0}$, $M_{1,1}$ 
and $M_{1,2}$, which contribute to the first order radiative energy loss.
For the interaction rate we find
\bea
\label{Gamma}
\Gamma (E) &=& \frac{1}{2 E} 2\,{\rm Im}\,M_\mathrm{tot} 
 = \frac{1}{2 E}\, (2\, {\rm Im}\,M_{1,0} 
                   +2\, {\rm Im}\,M_{1,1}
                   +2\, {\rm Im}\,M_{1,2}),
\eea
where (see Eqs.~(\ref{M10_final}), (\ref{M11_final}), and (\ref{M12_final}))
%\begin{widetext}
%
\beqar
2\,{\rm Im}\,M_{1,0} &=& 8 E \; g^4 T  \; [t_a, t_c] \,[t_c, t_a] \; 
\int \frac{d^3k} {(2 \pi)^3 2 \omega} \; \frac{d^2q}{(2 \pi)^2} \;
\frac{\mu^2}{\bq^2 (\bq^2{+}\mu^2)} \; 
\frac{\bk^2}{(\bk^2{+}\chi)^2} \; ,
\nonumber 
\\
2\,{\rm Im}\,M_{1,1}
 &=& 8 E \; g^4 T  \; [t_a,t_c]\, [t_c, t_a]  \;
  \int 
\frac{d^3k}{(2 \pi)^3 2 \omega} \; 
\frac{d^2q}{(2 \pi)^2} \; \frac{\mu^2}{\bq^2 (\bq^2{+}\mu^2)} \;
\frac{-2 \, \bk\cdot(\bk{+}\bq)}{(\bk^2{+}\chi)\,((\bk{+}\bq)^2+\chi)} \;,
\nonumber 
\\
2\,{\rm Im}\,M_{1,2} &=& 8 E \; g^4 T  \; [t_a, t_c] \, [t_c, t_a] \;
\int \frac{d^3k}{(2 \pi)^3 2 \omega} \; 
\frac{d^2q}{(2 \pi)^2} \; \frac{\mu^2}{\bq^2 (\bq^2{+}\mu^2)} \; 
\frac{(\bk{+}\bq)^2}{((\bk{+}\bq)^2+\chi)^2}\; .
\eeqar{M1}
%
%\end{widetext}
%
Here $[t_a,t_c]$ is a color commutator. $m_g^2=\mu^2/2$ is the effective 
mass for gluons with hard momenta $k\gtrsim T$, and $\chi\equiv M^2 x^2 
+ m_g^2$ where $x$ is the longitudinal momentum fraction of the heavy quark 
carried away by the emitted gluon. We assume constant coupling $g$.

By using the above equations, the interaction rate becomes
%
%\begin{widetext}
\beqar
\Gamma(E)
&=& 4 \; g^4 T  \; [t_a,t_c] \, [t_c, t_a] \;
\int \frac{d^3k}{(2 \pi)^3 2 \omega} \; \frac{d^2q}{(2 \pi)^2} \; 
\frac{\mu^2}{\bq^2 (\bq^2{+}\mu^2)} \; 
\left( \frac{\bk}{\bk^2+\chi}-\frac{\bk+\bq}{(\bk{+}\bq)^2+\chi}\right)^2 
\nonumber \\
&\approx& D_R \frac{C_R \alpha_s}{\pi}\,C_2(G) \alpha_s T 
\int \frac{dx}{x}\, \frac{d^2k}{\pi}\, \frac{d^2q}{\pi} \, 
\frac{\mu^2}{\bq^2 (\bq^2+\mu^2)}  
\left( \frac{\bk}{\bk^2+\chi} - \frac{\bk+\bq}{(\bk{+}\bq)^2+\chi}\right)^2\;,
\eeqar{Gamma1}
%\end{widetext}
%
where we used $[t_a,t_c] \, [t_c, t_a] = C_2 (G) C_R D_R$ (with $C_2(G)=3$, 
$C_R=\frac{4}{3}$, and $D_R=3$) and, in the second step, the soft 
rescattering approximation $|\bk|\ll k_z\approx \omega$.

The interaction rate sums over all initial and final colors of the heavy 
quark. The heavy quark radiative energy loss per unit length is obtained 
from the above expression for the interaction rate by weighting it with
the energy $\omega$ of the emitted gluon and averaging over the initial 
color of the heavy quark \cite{BT,BT_fermions,Le_Bellac}:
\beqar
\frac{d E_{\mathrm{dyn}}}{dL}
&=& \frac{1}{D_R}  \int d\omega\, \omega \frac{d\Gamma(E)}{d\omega} 
\approx \frac{E}{D_R} \int dx\, x\frac{d\Gamma(E)}{dx} .\quad
\eeqar{dEdl}
This leads to 
%
%\begin{widetext}
\beqar
\frac{\Delta E_{\mathrm{dyn}}}{E} 
&=& \frac{C_R \alpha_s}{\pi}\,\frac{L}{\lambda_\mathrm{dyn}}  
    \int dx \,\frac{d^2k}{\pi} \,\frac{d^2q}{\pi} \, 
    \frac{\mu^2}{\bq^2 (\bq^2+\mu^2)}
    \left( \frac{\bk}{\bk^2+\chi} - \frac{\bk{+}\bq}{(\bk{+}\bq)^2+\chi}
    \right)^2
\label{DeltaEDyn}
\\
&=& \frac{C_R \alpha_s}{\pi}\,\frac{L}{\lambda_\mathrm{dyn}}  
    \int dx\, d \bk^2 \, d \bq^2 \, 
\frac{\mu^2}{\bq^2 \, (\bq^2+\mu^2)} \; \frac{1}{\bk^2+\chi} \;
\nonumber \\
&\times& 
\left[ 
       \frac{\bq^2+\chi}{\sqrt{\chi^2 + 2\chi(\bk^2{+}\bq^2)
                                              + (\bk^2{-}\bq^2)^2}}
           - \frac{\chi}{\bk^2+\chi} 
     + \frac{\bq^2 \, \chi\,(\chi - 3\bk^2 + \bq^2)}
            {\left[\chi^2 + 2\chi(\bk^2{+}\bq^2) 
                          + (\bk^2{-}\bq^2)^2\right]^{\frac{3}{2}}}
\right] ,
\eeqar{DeltaEfi}
%\end{widetext}
%
where the second step is obtained after angular integration and we defined 
a ``dynamical mean free path'' $\lambda_\mathrm{dyn}$ through
\beq
\lambda_\mathrm{dyn}^{-1} \equiv C_2(G) \alpha_s T \; = 3 \alpha_s T\,.
\eeq{lambda_dyn}
Under the assumption that $\alpha_s$ is not running, Eq.~(\ref{DeltaEfi}) can 
be further analytically integrated over $0\leq |\bk| \leq \km$ 
where $\km= 2 E \sqrt{x (1-x)}$~\cite{DG_Ind}. We obtain 
\beqar
\frac{\Delta E_{\mathrm{dyn}}}{E} = 
\frac{C_R \alpha_s}{\pi^2}\,\frac{L}{\lambda_\mathrm{dyn}}
\int dx \, d^2q\, \mathcal{J}_{\mathrm{dyn}}(\bq,x)\,,
\eeqar{DeltaEq}
where
%
%\begin{widetext}
%
\bea
\mathcal{J}_{\mathrm{dyn}}(\bq,x) &=& \frac{\mu^2}{2 \, \bq^2 \, (\bq^2+\mu^2)} 
\Biggl[ -1 -\frac{2\,\km^2}{\km^2+\chi} 
           +\frac{\bq^2-\km^2+\chi}
                 {\sqrt{\bq^4+2\bq^2(\chi{-}\km^2) +(\km^2{+}\chi)^2}}
\label{JdynQ}
\\ 
&+&
\frac{2 \, (\bq^2+2\chi)}{\bq^2\sqrt{1{+}\frac{4\chi}{\bq^2}}}
\ln\Biggl(\frac{\km^2{+}\chi}{\chi}\,
         \frac{(\bq^2{+}3\chi) + \sqrt{1{+}\frac{4\chi}{\bq^2}}\,
              (\bq^2{+}\chi)}
              {(\bq^2{-}\km^2{+}3\chi) + \sqrt{1{+}\frac{4\chi}{\bq^2}}
               \sqrt{\bq^4+2\bq^2(\chi{-}\km^2)+(\km^2{+}\chi)^2}}
   \Biggr)\Biggr] .
\nonumber
\eea
%
%\end{widetext}
%

Alternatively, Eq.~(\ref{DeltaEfi}) can be integrated over 
$0\leq |\bq| \leq \qm$ where $\qm= \sqrt{4 E T}$~\cite{Adil}. 
This leads to an analytical expression for radiated gluon spectrum 
$\tilde{\mathcal{J}}_{\mathrm{dyn}}(\bk,x)$. We find
\beqar
\frac{\Delta E_{\mathrm{dyn}}}{E} = 
\frac{C_R \alpha_s}{\pi^2}\,\frac{L}{\lambda_\mathrm{dyn}}
\int dx \, d^2k\, \tilde{\mathcal{J}}_{\mathrm{dyn}}(\bk,x)\,,
\eeqar{DeltaEk}
where
%
%\begin{widetext}
\beqar
\tilde{\mathcal {J}}_{\mathrm{dyn}} (\bk,x) &=&  
\frac{2\,\chi\,\mu^2}{(\bk^2{+}\chi) \, {\cal G}(\chi,\bk,-\mu^2 )}   
\left(1-\frac{\bk^2{+}\chi}{\sqrt{{\cal G}(\chi,\bk,\qm^2)}}\right)
\nonumber \\
&-& \frac{\mu^2\,(\bk^2{+}\mu^2{-}3\chi)}
         {2\,(\bk^2{+}\chi)\,{\cal G}(\chi,\bk,-\mu^2)}
\left( \frac{\bk^2{-}3\chi}{\bk^2{+}\chi} - 
       \frac{\bk^2{-}\qm^2{-}3\chi}{\sqrt{{\cal G}(\chi,\bk,\qm^2)}} 
\right)
\nonumber \\ 
&-& \left( \frac{\chi} {(\bk^2{+}\chi)^2} 
     - \frac{2 \chi-\mu^2}{(\bk^2{+}\chi)\sqrt{{\cal G}(\chi,\bk,-\mu^2)}}
     + \frac{\chi\,(\bk^2{+}\chi{-}\mu^2)}
            {{\cal G}(\chi,\bk,-\mu^2)^{\frac{3}{2}}}
\right) 
\ln\left(\frac{\mu^2}{\qm^2{+}\mu^2}\right)
\nonumber \\
&-&\frac{\chi}{(\bk^2{+}\chi)^2}  
\ln\left[\frac{(\bk^2{+}\chi)^2 - (\bk^2{-}\chi)\,\qm^2 + 
               (\bk^2{+}\chi)\sqrt{{\cal G}(\chi,\bk,\qm^2)}}
              {2\,(\bk^2{+}\chi)^2}
   \right] 
\nonumber \\
&+& \frac{1}{\sqrt{{\cal G}(\chi,\bk,-\mu^2)}}
\left( \frac{2\chi{-}\mu^2}{\bk^2{+}\chi} 
     - \frac{\chi\,(\chi{+}\bk^2{-}\mu^2)}{{\cal G}(\chi,\bk,-\mu^2)}
\right) 
\nonumber\\
&& \times\ln\left[
\frac{(\bk^2{+}\chi)^2 - (\bk^2{+}\mu^2{-}\chi)\,\qm^2 + (\bk^2{-}\chi)\mu^2 
    + \sqrt{{\cal G}(\chi, \bk, \qm^2 )\, {\cal G}(\chi,\bk,-\mu^2)}}
     {(\bk^2{+}\chi)^2 + (\bk^2{-}\chi)\mu^2 
                       + (\bk^2{+}\chi)\sqrt{{\cal G}(\chi,\bk,-\mu^2)}}
\right] ,
\eeqar{JdynK} 
%\end{widetext}
%
with
\beq
{\cal G}(\chi,\bk,\kappa^2) \equiv 
(\bk^2{+}\chi)^2 - 2\kappa^2(\bk^2{-}\chi) + \kappa^4
\eeq{calG}
It is worth noting that each of the three contributions in Eq.~(\ref{M1}) 
diverges logarithmically in the limit of zero transverse momentum of the 
exchanged gluon, $\bq{\,\to\,}0$. The reason is that in a dynamical 
QCD medium both transverse and longitudinal gluon exchange contribute 
to the radiative energy loss. While Debye screening renders the 
longitudinal gluon exchange infrared finite, transverse gluon exchange
causes a well-known logarithmic singularity \cite{Le_Bellac,Wang_Dyn}, 
due to the absence of a magnetic screening \cite{fn1}. However, the 
infrared divergences cancel in the sum (\ref{Gamma}), giving rise to a 
finite energy loss rate. (This can be seen from Eq.~(\ref{JdynK}), where 
analytical integration over $\bq$ is performed.) This is a nontrivial
result since it was previously believed that in a dynamical medium a 
magnetic cutoff must be artificially introduced in order to avoid divergent 
results \cite{Le_Bellac,Wang_Dyn}. 

%%%%%%%%%%%%%%%%%%%%%%%%%%%%%%%%%%%%%%%%%%%%%%%%%%%%%%%%%%%%%%%%%%%%%%%%
\section{Radiative energy loss in a static QCD medium} 
\label{sec3}
%%%%%%%%%%%%%%%%%%%%%%%%%%%%%%%%%%%%%%%%%%%%%%%%%%%%%%%%%%%%%%%%%%%%%%%%

Let us now briefly revisit for comparison the heavy quark radiative energy 
loss in a static QCD medium, using a derivation which parallels that of the 
previous section and thus clearly exhibits the differences between the two 
situations. We again consider an on-shell heavy quark produced in the remote 
past propagating through an infinite QCD medium that now consists of randomly 
distributed static scattering centers. The static interactions are modeled 
as Gyulassy-Wang \cite{Gyulassy_Wang} color-screened Yukawa potentials
\beqar 
V_n &=& V(q_n)\, e^{i q_n{\cdot}x_n}
\label{gwmod}\\
&=& 2\pi\,\delta(q^0)\,v(\vq_n)\, 
e^{-i \vq_n\cdot\vx_n} \; T_{a_n}(R)\otimes T_{a_n}(n) \; ,
\nonumber
\eeqar
where $\vx_n$ is the location of the $n^\mathrm{th}$ scattering center,
the two $T$ symbols (with $a_n$ being summed over) denote the color 
matrices of the heavy quark and the $n^\mathrm{th}$ scattering center, 
and $v(\vq_n)\equiv {4\pi\alpha_s}/(\vq_n^2+\mu^2)$. 

\begin{figure}[ht]
\vspace*{4cm} 
\includegraphics{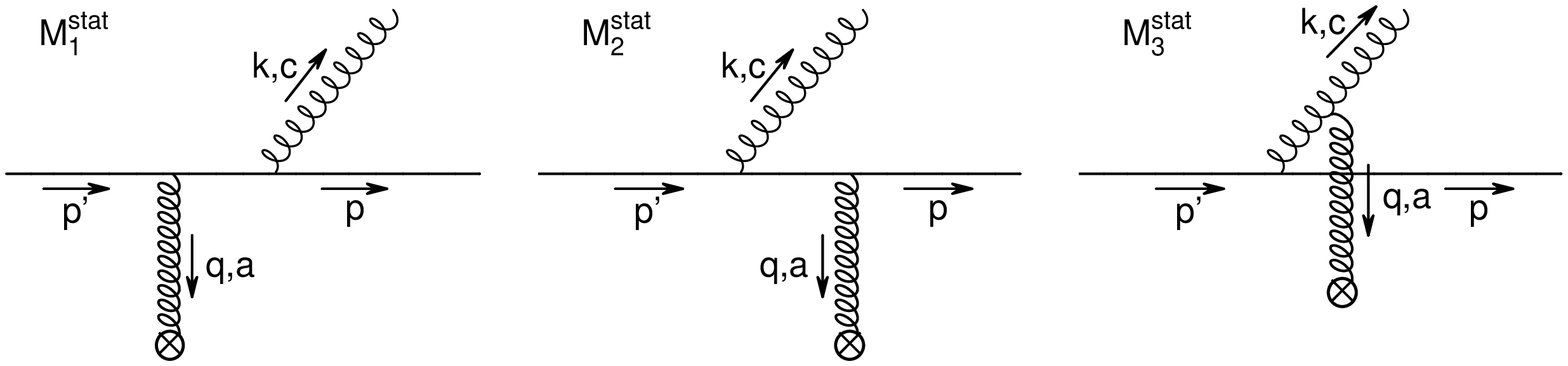}
\caption{Feynman diagrams $M^\mathrm{stat}_{1}$, $M^\mathrm{stat}_{2}$ and 
$M^\mathrm{stat}_{3}$ contributing to the soft gluon radiation amplitude in
a static medium to first order in opacity. The static color charge has color 
$a$ and exchanges momentum $q=(0,q_z,\bq)$ with the heavy quark. The 
radiated gluon has color $c$ and carries momentum $k=(\omega, k_z,\bk)$.
\label{staticFO}}
\end{figure}

The diagrams contributing to the radiative energy loss at first order in 
opacity are shown in Fig.~\ref{staticFO}. As seen in the figure, the quark 
scatters of one of the color centers with momentum $q=(0,q_z,\bq)$ and 
radiates a gluon with momentum $k=(\omega, k_z,\bk)$. As in the previous 
section, energy-momentum conservation requires $p'=p+k+q$.  
%
%%%%%%%%%%%%%%%%%%%%%%%%%%%%%%% Figure 1 %%%%%%%%%%%%%%%%%%%%%%%%%%%%%%%%%%%%
%\begin{widetext}
%\end{widetext}
%%%%%%%%%%%%%%%%%%%%%%%%%%%%%%%%%%%%%%%%%%%%%%%%%%%%%%%%%%%%%%%%%%%%%%%%%%%%%
%

The procedure for the calculation of the diagrams shown in Fig.~1
%Fig.~\ref{staticFO} 
was already presented in~\cite{DG_Ind} (see in particular appendices A 
and B there), so we will not repeat it here. We obtain 
%
%\begin{widetext}
\beqar
M^\mathrm{stat}_{1}(p,k)&\approx& 
4 i g \, E \int \frac{d^4 q_1}{(2 \pi)^4} \; V(q_1)\, 
e^{i q_1\cdot x_1} \frac{p\cdot\epsilon}{(p{+}k)^2-M^2}\,t_c t_{a_1}\, , 
\nonumber \\
M^\mathrm{stat}_{2}(p,k)&\approx& 
4 i g \, E  \int \frac{d^4 q_1}{(2 \pi)^4} \;V(q_1)\, 
e^{i q_1\cdot x_1} \frac{{p'}\cdot\epsilon}{(p'-k)^2-M^2}\,t_{a_1} t_c\, , 
\nonumber \\
M^\mathrm{stat}_{3}(p,k)&\approx& 
4i g \, E \int \frac{d^4 q_1}{(2 \pi)^4} \; V(q_1)\, e^{i q_1\cdot x_1} 
\frac{\bm{\epsilon}\cdot(\bk{+}{\bq}_1)}
{(k+q_1)^2-m_g^2} \, [t_c, t_{a_1}]\, ,
\eeqar{M123_stat}
%\end{widetext}
%
where $\epsilon\equiv\epsilon(k)=\left(0,\,\frac{2\bm{\epsilon}\cdot\bk}
{x(k_z{+}\omega)},\bm{\epsilon}\right)$ (with $\bm{\epsilon}=(1,0)$ or (0,1)) 
is the polarization vector of the emitted gluon. By using Eqs.~(\ref{gwmod}),
(\ref{ppprimek}), and (\ref{k+q}), together with $p\cdot\epsilon \approx 
{p'}\cdot\epsilon \approx \bm{\epsilon}\cdot\bk/x$ \cite{GLV,DG_Ind}, 
Eq.~(\ref{M123_stat}) becomes
%
%\begin{widetext}
\beqar  
M^\mathrm{stat}_{1}&\approx& 4 i g \, E \int  \frac{d^3q_1}{(2 \pi)^3 }\, 
v(\vq_1)\, e^{-i \vq_1\cdot\vx_1} \;
\frac{\bm{\epsilon}\cdot\bk}{\bk^2+\chi} 
\, t_c t_{a_1}\, T_{a_1} , 
\nonumber \\
M^\mathrm{stat}_{2}&\approx& - 4 i g \, E  \int  \frac{d^3q_1}{(2 \pi)^3}\, 
v(\vq_1)\, e^{-i \vq_1\cdot\vx_1} \;
\frac{\bm{\epsilon}\cdot\bk}{\bk^2+\chi} \,
t_{a_1} t_c \, T_{a_1} \; , 
\nonumber \\
M^\mathrm{stat}_{3}&\approx& - 4i g \, E  \int  \frac{d^3q_1}{(2 \pi)^3}\, 
v(\vq_1)\,e^{-i \vq_1\cdot\vx_1} \;
\frac{\bm{\epsilon}\cdot(\bk{+}\bq_1)}{(\bk{+}\bq_1)^2+\chi} 
\, [t_c, t_{a_1}]\,  T_{a_1}  \;.
\eeqar{M123_stat1}
%\end{widetext}
%
Therefore
\beqar  
M^\mathrm{stat}_\mathrm{tot}(p,k) &=&
M^\mathrm{stat}_{1}(p,k)+M^\mathrm{stat}_{2}(p,k)+M^\mathrm{stat}_{3}(p,k)
\nonumber \\
&\approx& 
4 i g \, E\, [ t_c, t_{a_1}] \, T_{a_1} \int \frac{d^3q_1}{(2 \pi)^3 }\,
v(\vq_1)\,e^{-i \vq_1\cdot\vx_1} 
\left[\frac{\bm{\epsilon}\cdot\bk}{\bk^2+\chi} - 
      \frac{\bm{\epsilon}\cdot(\bk{+}\bq_1)}{(\bk{+}\bq_1)^2+\chi} \right]
\;.
\eeqar{Mtot_stat}
Squaring this and ensemble-averaging the result over the positions $\vx_1$ 
of the scattering centers gives
\beqar  
\langle|M^\mathrm{stat}_\mathrm{tot}|^2\rangle(p,k) &\approx&
16 g^2 E^2 [ t_c, t_{a_1}][ t_{a_2}, t_c] \, T_{a_1} T_{a_2} 
\int \frac{d^3q_1}{(2\pi)^3}\; \frac{d^3q_2}{(2\pi)^3}\;
\langle e^{-i (\vq_1-\vq_2)\cdot\vx_1}\rangle v(\vq_1)\,v(\vq_2)\, 
\nonumber \\
&&\times \sum_{\bm{\epsilon}}
\left(\frac{\bm{\epsilon}\cdot\bk}{\bk^2+\chi} -
      \frac{\bm{\epsilon}\cdot(\bk{+}\bq_1)}{(\bk{+}\bq_1)^2+\chi} \right)
\left(\frac{\bm{\epsilon}\cdot\bk}{\bk^2+\chi} -
      \frac{\bm{\epsilon}\cdot(\bk{+}\bq_2)}{(\bk{+}\bq_2)^2+\chi} \right) 
\nonumber \\
&\approx& 16 g^2 E^2\, [ t_c, t_a][ t_a, t_c] \, 
C_2(T) \frac{d_T}{d_g} \frac{1}{V} \int \frac{d^3q}{(2 \pi)^3 }\; |v(\vq)|^2  
\left(\frac{\bk}{\bk^2+\chi} - \frac{\bk{+}\bq}{(\bk{+}\bq)^2+\chi}\right)^2 
\;, 
\eeqar{Mtot_stat2}
In the last step we used \cite{GLV} $\tr(T_{a_1} T_{a_2})= \delta_{a_1 a_2}
C_2 (T) d_T/d_g$ (we assume that all scattering centers (``target partons'')
have the same $d_T$-dimensional color representation with Casimir $C_2(T )$). 
Furthermore, we assumed (similarly to \cite{GLV}) that the ensemble average 
over the phase factor gives
\beqar
\langle e^{-i(\vq_1-\vq_2)\cdot\vx}\rangle = \frac{(2\pi)^3}{V} 
\delta^{(3)}(\vq_1{-}\vq_2)\,,
\eeqar{bave}
where $V=L A_\perp$, with $L$ and $A_\perp$ being the depth (along the 
heavy quark's path) and transverse area of the medium.

We can now compute the interaction rate for the heavy quark in a static 
medium of scatterers with color representation $T$:
\beqar
\Gamma_T^\mathrm{stat}(E)&=&
\int \frac{d^3p}{(2\pi)^3 2E}\; \frac{d^3k}{(2\pi)^3 2\omega}\; 
(2\pi)^4 \delta^{(4)}(p'-p-k-q)\,\frac{1}{2E}\, 
\langle|M^\mathrm{stat}_\mathrm{tot}|^2\rangle(p,k)
\nonumber\\
&=& \int \frac{d^3k}{(2\pi)^3 2\omega}\,
2\pi\delta\Bigl(q_z-\frac{\bk^2{+}M^2x^2{+}m_g^2}{2xE}\Bigr)\,
\frac{1}{4E^2}\, \langle|M^\mathrm{stat}_\mathrm{tot}|^2\rangle
\nonumber\\
&\approx& 4 \alpha_s [t_c, t_a] [t_a, t_c] \, C_2 (T)\, \frac{d_T}{d_g}  
\frac{1}{V} \int \frac{dx}{x}\,\frac{d^2k}{(2\pi)^2}\,\frac{d^2q}{(2\pi)^2}\,
|v(0,\bq)|^2  \left(\frac{\bk}{\bk^2+\chi} 
                  - \frac{\bk{+}\bq}{(\bk+\bq)^2+\chi}\right)^2
\nonumber\\
&=& D_R d_T \frac{C_R \alpha_s}{\pi} \frac{1}{\lambda_{T}}  
\int \frac{dx}{x}\, \frac{d^2k}{\pi}\,\frac{d^2q}{\pi}\, 
\frac{\mu^2}{(\bq^2+\mu^2)^2} 
\left(\frac{\bk}{\bk^2+\chi} - \frac{\bk{+}\bq}{(\bk{+}\bq)^2+\chi}\right)^2
\;,
\eeqar{GammaStat}
where we defined the mean free path for a heavy quark
scattering off quark-type (``$q$'') or gluon-type (``$g$'') scattering 
centers through \cite{WHDG,Gyulassy_Wang,GLV} 
\beqar
\frac{1}{\lambda_{T}}=\frac{3}{8}C_2(T)
\frac{1}{V_T} \int \frac{d^2q}{(2\pi)^2}\,
|v(0,\bq)|^2\; 
&\Longrightarrow& \left\{ \begin{array}{ll}
% \hspace{0.15in}
\frac{1}{\lambda_{g}}=\frac{9}{4}\cdot2\pi\frac{\alpha^2_s}{\mu^2}\cdot\rho_g 
= \frac{9 \pi}{2} \frac{\alpha^2_s}{\mu^2}\,\frac{1.202}{\pi^2}\cdot16T^3
\\[2ex] 
\frac{1}{\lambda_{q}}=2 \pi \frac{\alpha^2_s}{\mu^2}\cdot \rho_q =
2 \pi \frac{\alpha^2_s}{\mu^2}\,\frac{1.202}{\pi^2}\cdot9 n_f T^3 
\end{array}\right.
\eeqar{lambdaT}
After averaging over the initial colors of the heavy quark and the scattering
centers and weighting the rate with the energy $\omega$ of the emitted gluon, 
the heavy quark energy loss in an infinite static QCD medium is then given by
\beqar
\frac{dE_{\rm{stat}}}{dL}=\frac{1}{D_R} \int d\omega\, \omega 
\left(\frac{1}{d_g}\frac{d\Gamma^\mathrm{stat}_g(E)}{d\omega} 
    + \frac{1}{d_q}\frac{d\Gamma^\mathrm{stat}_q(E)}{d\omega}\right).
\eeqar{dEstatdL}
This leads to 
\beqar
\frac{\Delta E_{\rm{stat}}}{E} = \frac{C_R \alpha_s}{\pi} 
\frac{L}{\lambda_\mathrm{stat}} \int dx\, \frac{d^2k}{\pi}\,
\frac{d^2q}{\pi}\; \frac{\mu^2}{(\bq^2+\mu^2)^2}
\left(\frac{\bk}{\bk^2+\chi} - \frac{\bk{+}\bq}{(\bk{+}\bq)^2+\chi}\right)^2,
\eeqar{DeltaEStat}
with
\beqar
\frac{1}{\lambda_\mathrm{stat}}=\frac{1}{\lambda_{g}}+\frac{1}{\lambda_{q}}=
6 \frac{1.202}{\pi^2} \frac{1+\frac{n_f}{4}}{1+\frac{n_f}{6}} 3 \alpha_s T
= c(n_f) \; \frac{1}{\lambda_\mathrm{dyn}} \;,
\eeqar{lambda_stat}
where $c(n_f) \equiv 6 \frac{1.202}{\pi^2} \frac{1+n_f/4}{1+n_f/6}$
is a slowly increasing function of $n_f$ that varies between 
$c(0)\approx 0.73$ and $c(\infty)\approx1.09$. For a typical value 
$n_f=2.5$ (which we use in this paper), $c(2.5) \approx 0.84\simeq1$.

As in the previous section, under the assumption that $\alpha_S$ is not 
running, Eq.~(\ref{DeltaEStat}) can be analytically integrated over 
$\bk$ or $\bq$. Integration over $\bk$ yields
\beqar
\frac{\Delta E_{\rm{stat}}}{E} = 
\frac{C_R \alpha_S}{\pi^2} \;  \frac{L}{\lambda_{\rm{stat}}}  \int 
d x \;  d^2q \; \mathcal {J}_{\rm{stat}} (\bq,x)
\eeqar{DeltaEstatq}
with the simple relationship
\beqar
\mathcal{J}_{\rm{stat}} (\bq,x)&=&\mathcal{J}_{\mathrm{dyn}} (\bq,x) \;
\frac{\bq^2}{\bq^2+\mu^2} 
\eeqar{JstatQ}
where $\mathcal{J}_{\mathrm{dyn}}(\bq)$ is given by Eq.~(\ref{JdynQ}).
Integrating Eq.~(\ref{DeltaEStat}) instead over $\bq$ gives
\beqar
\frac{\Delta E_{\mathrm{stat}}}{E} = 
\frac{C_R \alpha_S}{\pi^2} \;  \frac{L}{\lambda_\mathrm{stat}}  
\int dx\,d^2k\, \tilde{\mathcal{J}}_{\mathrm{stat}}(\bk,x)\; ,
\eeqar{DeltaEstatk}
where
%\begin{widetext}
\beqar
\tilde{\mathcal{J}}_{\mathrm{stat}}(\bk,x) &=& 
-\frac{\chi\,\qm^2}{(\qm^2{+}\mu^2)(\bk^2{+}\chi)^2}
+\frac{\,2 \chi - 3 \mu^2}{2\,{\cal G}(\chi,\bk,-\mu^2)}
+\frac{3\chi\mu^2(3\bk^2{+}\mu^2{-}\chi)}{{\cal G}(\chi,\bk,-\mu^2)^2}
\nonumber\\  
&+& \frac{\mu^2\sqrt{{\cal G}(\chi,\bk,\qm^2)}}
         {(\qm^2{+}\mu^2)\,{\cal G}(\chi,\bk,-\mu^2)}
\left(\frac{\mu^2{-}2\chi}{\bk^2{+}\chi} 
    + \frac{3\chi (\bk^2{+}\chi{-}\mu^2)}{{\cal G}(\chi,\bk,-\mu^2)} 
    - \frac{2\chi (\bk^2{+}\chi{+}\qm^2)}{{\cal G}(\chi,\bk,\qm^2)} 
\right)
\nonumber \\
&+& \frac{\mu^2 (\bk^2{+}\chi{-}\qm^2)}
         {2{\cal G}(\chi,\bk,-\mu^2)\sqrt{{\cal G}(\chi,\bk,\qm^2 )}}
\nonumber\\ 
&+& \frac{\mu^2}{{\cal G}(\chi,\bk,-\mu^2)^{\frac{3}{2}}}
\left( - \frac{4\chi \bk^2}{\bk^2 +\chi} 
       + (\bk^2{+}\chi{+}\mu^2)
         \left(\frac{6\chi\bk^2}{{\cal G}(\chi,\bk,-\mu^2)} - 1\right)
\right) 
\nonumber \\
&& \times\ln\left[\frac{\mu^2}{\qm^2{+}\mu^2}
\frac{(\bk^2{+}\chi)^2 - (\bk^2{+}\mu^2{-}\chi)\qm^2 + (\bk^2{-}\chi)\mu^2 
     + \sqrt{{\cal G}(\chi,\bk,\qm^2) {\cal G}(\chi,\bk,-\mu^2)}}
     {(\bk^2{+}\chi)^2 + (\bk^2{-}\chi)\mu^2 
                       + (\bk^2{+}\chi)\sqrt{{\cal G}(\chi,\bk,-\mu^2)}}
            \right] , 
\eeqar{JstatK}
%\end{widetext}
with ${\cal G}$ given by Eq.~(\ref{calG}). Note that the emitted gluon 
spectra, Eq.~(\ref{JdynK}) for $\tilde{\mathcal{J}}_{\mathrm{dyn}}(\bk,x)$ 
and Eq.~(\ref{JstatK}) for $\tilde{\mathcal{J}}_{\mathrm{stat}}(\bk,x)$, 
while showing some similarities, don't exhibit a similarly simple analytical 
relationship as was the case for $\mathcal{J}_{\mathrm{dyn}}(\bq,x)$ and 
$\mathcal{J}_{\mathrm{stat}}(\bq,x)$ (see Eq.~(\ref{JstatQ})).  

Finally, we can compare the radiative energy loss rates to first order in
opacity in dynamic and static QCD media, Eqs.~(\ref{DeltaEDyn}) and 
(\ref{DeltaEStat}). Both expressions yield an energy loss that increases
linearly with the path length $L$ through the medium. This reflects our
neglect of LPM interference effects \cite{MVWZ:2004,BDMS} -- our result 
corresponds to the Bethe-Heitler limit. In spite of this and many other 
similarities between Eqs.~(\ref{DeltaEDyn}) and (\ref{DeltaEStat}), they 
feature two important differences: The first is an $\mathcal{O}(15\%)$ 
decrease in the mean free path
\beq
\lambda_\mathrm{dyn} \Longleftrightarrow 
\lambda_\mathrm{stat}=\frac{\lambda_\mathrm{dyn}}{c(n_f)}
\eeq{diff_lam}
which increases the energy loss rate in the dynamical medium by 
$\mathcal{O}(20\%)$. The second difference is a change in the shape and 
normalization of the emitted gluon spectrum, which in the energy loss rate 
is reflected by the replacement
\beq
\left[ \frac{\mu^2}{\bq^2 (\bq^2{+}\mu^2)} \right]_\mathrm{dyn} 
\Longleftrightarrow
\left[\frac{\mu^2}{(\bq^2{+}\mu^2)^2}\right]_\mathrm{stat}.
\eeq{diff_spec}
As we will see in the next section, this second difference leads to an 
additional significant increase of the heavy quark energy loss rate and 
of the emitted gluon radiation spectrum by about $50\%$ for the dynamical 
QCD medium.

%%%%%%%%%%%%%%%%%%%%%%%%%%%%%%%%%%%%%%%%%%%%%%%%%%%%%%%%%%%%%%%%%%%%%%%%
\section{Numerical results}
\label{sec4}
%%%%%%%%%%%%%%%%%%%%%%%%%%%%%%%%%%%%%%%%%%%%%%%%%%%%%%%%%%%%%%%%%%%%%%%%

In this section we present numerical results for the total radiative 
energy loss, as well as its differential rate with respect to the energy
fraction and transverse momentum carried by the exchanged and emitted gluons,
to first order in opacity, using the expressions for infinite QCD media 
derived in the two preceding sections. Specifically, we discuss a static 
quark-gluon plasma of temperature $T{\,=\,}225$\,MeV, with $n_f{\,=\,}2.5$ 
effective light quark flavors and strong interaction strength 
$\alpha_s{\,=\,}0.3$, as representative of average conditions encountered 
in Au+Au collisions at RHIC. Further below we will raise the temperature of 
the medium to $T{\,=\,}400$\,MeV to simulate (average) conditions in Pb+Pb 
collisions at the LHC. For the light quarks we assume that their mass is 
dominated by the thermal mass $M{\,=\,}\mu/\sqrt{6}$, where 
$\mu{\,=\,}gT\sqrt{1{+}N_f/6}\approx 0.5$ GeV is the Debye screening 
mass. The charm mass is taken to be $M{\,=\,}1.2$\,GeV, and for 
the bottom mass we use $M{\,=\,}4.75$\,GeV. As noted before, the radiative
energy loss in the Bethe-Heitler limit considered here increases linearly
with the path length $L$ traveled by the fast quark -- for normalization
purposes we will set this path length to a standard value of $L=5$\,fm 
throughout.

%%%%%%%%%%%%%%%%%%%%%%%% Fig. 2 %%%%%%%%%%%%%%%%%%%%%%%%%%%%%%%%%%%%%%%%%%%%
\begin{figure}[ht]
\vspace*{5cm} 
\includegraphics{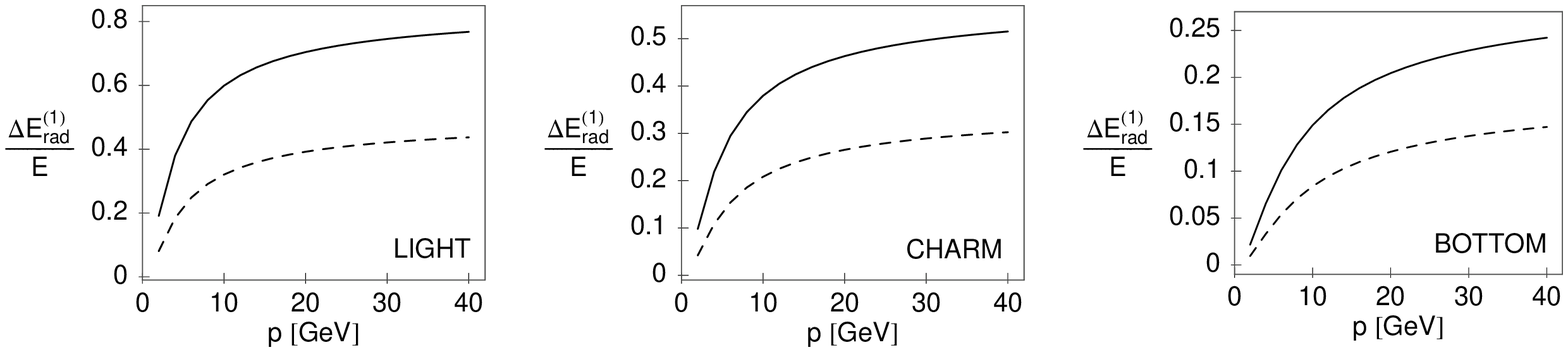}
\caption{Fractional radiative energy loss for an assumed path length 
$L=5$\,fm as a function of momentum for light, charm and bottom quarks 
(left, center, and right panels, respectively). Full and dashed curves 
correspond to the energy loss in a dynamical and a static QCD medium, 
respectively, and are obtained from Eqs.~(\ref{DeltaEDyn}) and 
Eq.~(\ref{DeltaEStat}).}
\label{DynVSStatE}
\end{figure}
%%%%%%%%%%%%%%%%%%%%%%%%%%%%%%%%%%%%%%%%%%%%%%%%%%%%%%%%%%%%%%%%%%%%%%%%%%%%

In Figure~\ref{DynVSStatE} we compare the momentum dependence of the 
radiative energy loss over an assumed path length $L=5$\,fm for a dynamical 
and a static QCD medium (as given by Eqs.~(\ref{DeltaEDyn}) 
and~(\ref{DeltaEStat})) . For all three types of quarks, the dynamical 
medium is seen to cause almost $70\%$ higher energy loss than the
static medium. The left panel in Fig.~\ref{ratio} below shows that
$\sim 50\%$ of this increase arises from the larger values of the 
function $\tilde{\mathcal{J}}_\mathrm{dyn}(\bk,x)$ which describes the 
shape of the emitted gluon spectrum, with an additional $\sim 20\%$ 
increase stemming from the shorter mean free path in the dynamical 
medium.

To better understand the kinematic distribution of the dynamical medium 
effects we will now study the energy loss differentially as a function of 
the gluon energy fraction $x=\omega/E$ and the transverse momenta $\bk$ 
and $\bq$ of the emitted and exchanged gluons. We define the 
double-differential energy loss spectra
\bea
\mathcal{S}(\bq,x) &=& \frac{1}{E}\,\frac{d(\Delta E)}{dx\,d^2q}
 = \frac{C_R\alpha_s}{\pi^2}\,\frac{L}{\lambda}\,\mathcal{J}(\bq,x),
\label{Sqx}
\\
\tilde{\mathcal{S}}(\bk,x) &=& \frac{1}{E}\,\frac{d(\Delta E)}{dx\,d^2k}
 = \frac{C_R\alpha_s}{\pi^2}\,\frac{L}{\lambda}\,\tilde{\mathcal{J}}(\bk,x)
\label{Skx}
\eea
as well as their single-differential analogues
\bea
\mathcal{S}(\bq) &=& \frac{1}{E}\,\frac{d(\Delta E)}{d^2q}
 = \frac{C_R\alpha_s}{\pi^2}\,\frac{L}{\lambda}
   \int dx\,\mathcal{J}(\bq,x),
\label{Sq}
\\
\tilde{\mathcal{S}}(\bk) &=& \frac{1}{E}\,\frac{d(\Delta E)}{d^2k}
 = \frac{C_R\alpha_s}{\pi^2}\,\frac{L}{\lambda}
   \int dx\,\tilde{\mathcal{J}}(\bk,x),
\label{Sk}
\\
\mathcal{S}(x) &=& \frac{1}{E}\,\frac{d(\Delta E)}{dx}
 = \frac{C_R\alpha_s}{\pi^2}\,\frac{L}{\lambda}
   \int d^2q\,\mathcal{J}(\bq,x)
 = \frac{C_R\alpha_s}{\pi^2}\,\frac{L}{\lambda}
   \int d^2k\,\tilde{\mathcal{J}}(\bk,x).
\label{Sx}
\eea

\begin{figure}[ht]
\vspace*{6.0cm} 
\includegraphics{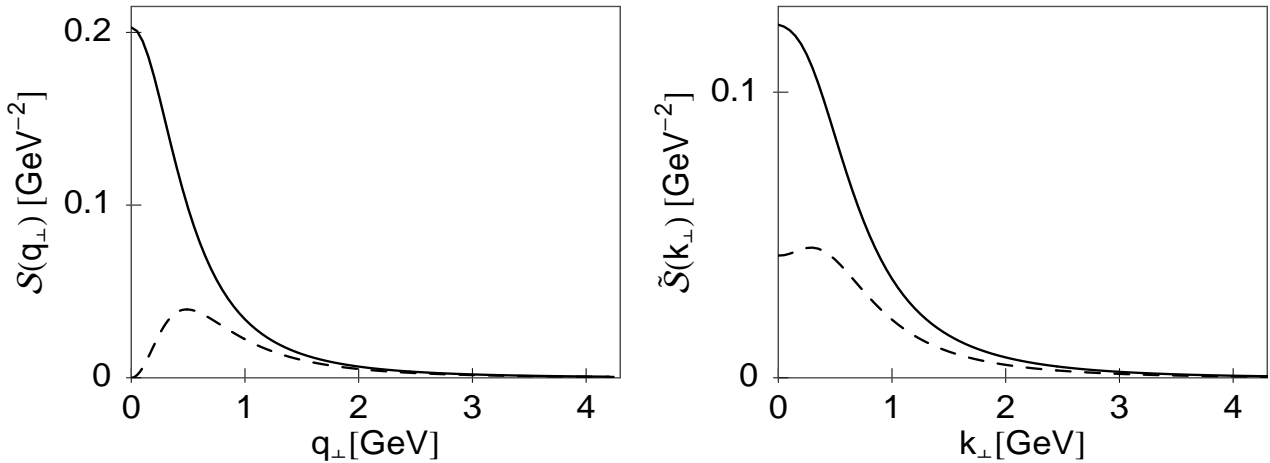}
\caption{Transverse momentum spectrum of the exchanged (Eq.~(\ref{Sq}), left) 
and emitted (Eq.~(\ref{Sk}), right) gluons for charm quarks traveling for 
$L=5$\,fm through a medium of dynamical (solid) or static (dashed) scatterers. 
%The spectrum 
%is integrated over $x$: $\mathcal{S}(\bq)= \int \mathcal{S}(\bq,x)\,dx$ and 
%$\tilde{\mathcal{S}}(\bk)= \int\tilde{\mathcal{S}}(\bk,x)\,dx$.
The initial charm quark energy was assumed to be 20 GeV. Note that 
$|\bq|$ and $|\bk|$ are denoted as q$_\perp$ and k$_\perp$ in the figure.}
\label{spectrum_qk}
\end{figure}

In Figure~\ref{spectrum_qk} we plot the transverse momentum dependence of 
the exchanged ($\mathcal{S}(\bq)$, left panel) and emitted gluon spectrum 
($\tilde{\mathcal{S}}(\bk)$, right panel). The largest differences between 
static (dashed) and dynamical media (solid) are observed at low transverse 
momenta $|\bq|,\,|\bk|\lesssim1$\,GeV, especially for the exchanged 
momentum spectrum $\mathcal{S}(\bq)$. The latter is seen to exhibit 
qualitatively different low-$\bq$ behavior in static and dynamical media, 
%
%%%%%%%%%%%%%%%%%%%%%%%% Fig. 3 %%%%%%%%%%%%%%%%%%%%%%%%%%%%%%%%%%%%%%%%%%%%
%%%%%%%%%%%%%%%%%%%%%%%%%%%%%%%%%%%%%%%%%%%%%%%%%%%%%%%%%%%%%%%%%%%%%%%%%%%%
%
but the difference is seen to rapidly disappear at $|\bq|>1$\,GeV (middle 
panel in Fig.~\ref{ratio}), as expected from Eq.~(\ref{diff_spec}). 

While the dynamical medium effects on the exchanged gluon spectrum are
mostly concentrated at low transverse momenta, the right panels in
Fig.~\ref{spectrum_qk} and (more clearly) Fig.~\ref{ratio} show that the 
effect on the {\em emitted} gluon spectrum extends over the entire 
transverse momentum region, causing enhancements by more than a factor 2.5 
at low $|\bk|$, and settling down at high $|\bk|\gtrsim1$\,GeV to an 
approximately $|\bk|$-independent enhancement by a factor $\sim 1.3$. 
%
%%%%%%%%%%%%%%%%%%%%%%%% Fig. 4 %%%%%%%%%%%%%%%%%%%%%%%%%%%%%%%%%%%%%%%%%%%%
\begin{figure}[ht]
\vspace*{5.2cm} 
\includegraphics{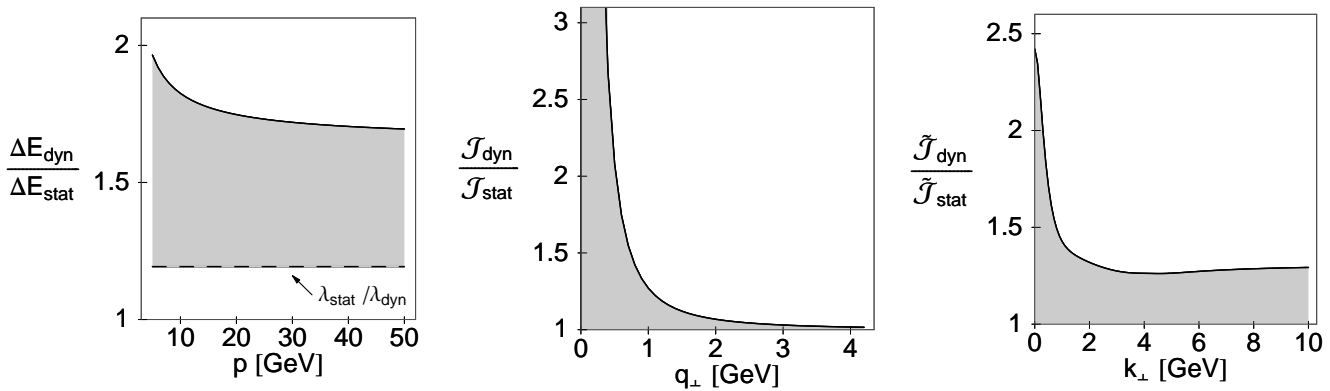}
\caption{{\sl Left panel:} Ratio of dynamical and static energy losses as 
a function of initial quark momentum $p$. The dashed line shows the $20\%$ 
increase in the energy loss due to the shorter mean free path 
$\lambda_\mathrm{dyn} < \lambda_\mathrm{stat}$ in the dynamical medium 
(see Eq.~(\ref{diff_lam})). The dashed region indicates the additional 
$\sim 50\%$ increase in the dynamical energy loss due to 
Eq.~(\ref{diff_spec}).
{\sl Middle panel:} $|\bq|$ dependence of the ratio of the energy loss 
integrands $\mathcal{J}(\bq)=\int dx\, \mathcal{J}(\bq,x)$ in 
Eqs.~(\ref{DeltaEq}), (\ref{DeltaEstatq}). 
{\sl Right panel:} $|\bk|$ dependence of the ratio of the energy loss 
integrands $\tilde{\mathcal{J}}(\bk)=\int dx\, \tilde{\mathcal{J}}(\bk,x)$ in 
Eqs.~(\ref{DeltaEk}), (\ref{DeltaEstatk}). Note that $|\bq|$ 
and $|\bk|$ are denoted in the figure as q$_\perp$ and k$_\perp$ respectively.}
\label{ratio}
\end{figure}
%%%%%%%%%%%%%%%%%%%%%%%%%%%%%%%%%%%%%%%%%%%%%%%%%%%%%%%%%%%%%%%%%%%%%%%%%%%%
%

\begin{figure}[ht]
\vspace*{6cm} 
\includegraphics{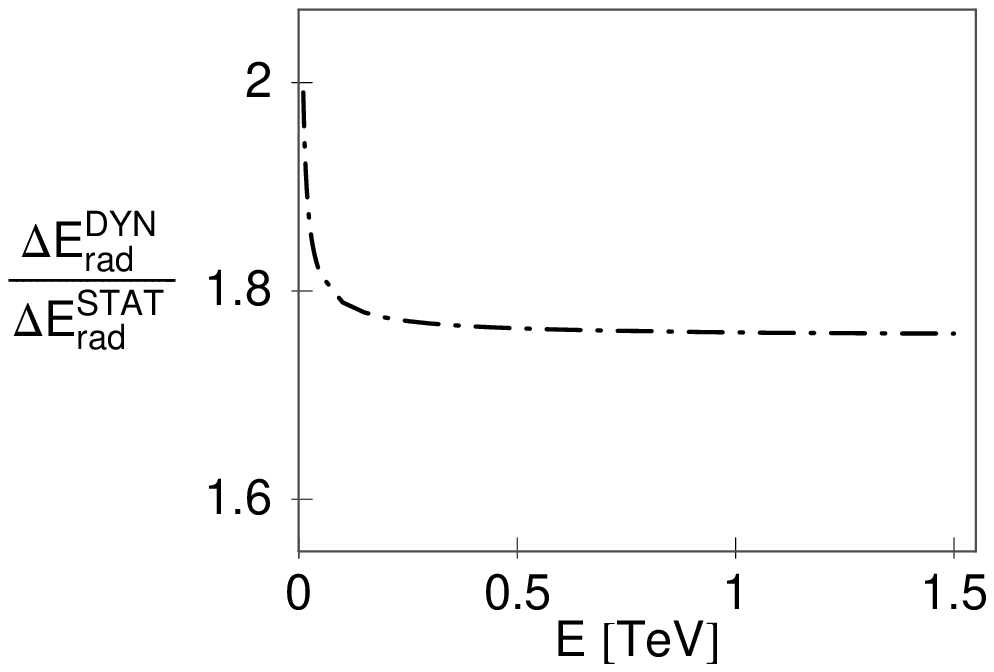}
\includegraphics{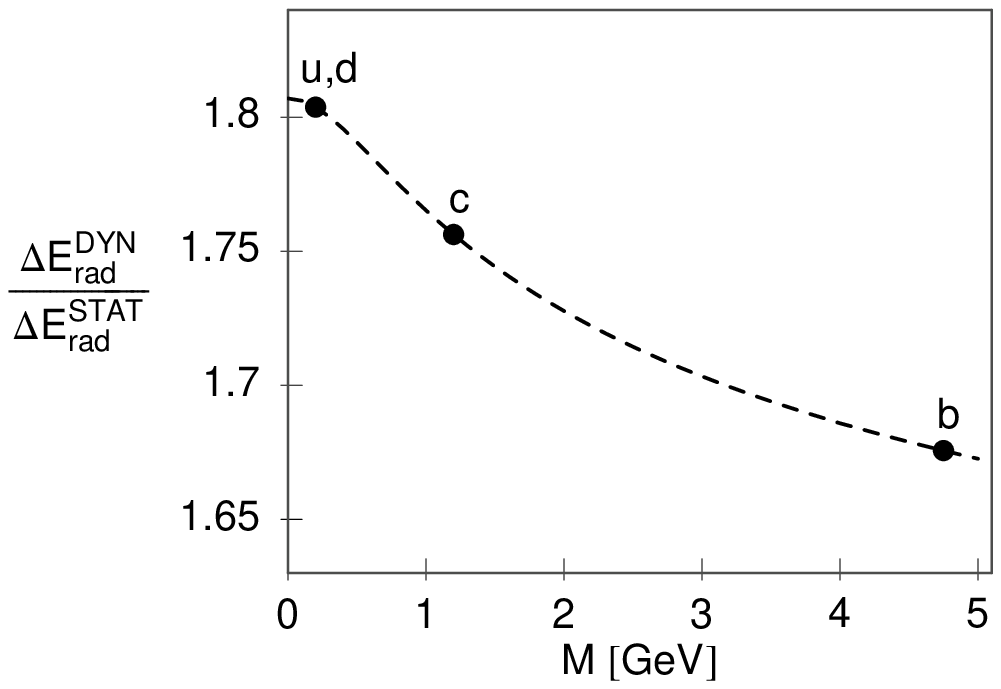}
\caption{Radiative energy loss in a medium of temperature $T=400$\,MeV
(``LHC conditions''). {\sl Left panel:} Ratio of the fractional radiative 
energy loss in dynamical and static media for charm quarks as a function 
of initial quark energy $E$. The ratio saturates quickly to a constant 
value above $E\sim 100$\,GeV.
{\sl Right panel:} Asymptotic value of the radiative energy loss ratio 
for high energy quarks as a function of their mass, with marks indicating
the light, charm and bottom quarks.}
\label{ratioLHC}
\end{figure}

The $\bk$-integrated effect on the radiative charm energy loss is shown 
in the left panel of Fig.~\ref{ratio}. The energy loss ratio between 
dynamical and static media is almost independent of the momentum $p$ of 
the fast charm quark, saturating at $\simeq 1.7$ above $p\gtrsim 20$\,GeV
and being even somewhat larger at smaller momenta. (This includes the 
$\sim20\%$ effect arising from the shorter mean free path in the 
dynamical medium.) We checked that the dynamical enhancement persists 
at constant level to the largest possible charm quark energies. This is 
shown explicitly in the left panel of Fig.~\ref{ratioLHC} where we plot 
the charm quark energy loss ratio for an otherwise identical medium of 
higher temperature $T=400$\,MeV (``LHC conditions''). There is no quark 
energy domain where the assumption of static scatterers in the medium 
becomes a valid approximation. 

%%%%%%%%%%%%%%%%%%%%%%%% Fig. 5 %%%%%%%%%%%%%%%%%%%%%%%%%%%%%%%%%%%%%%%%%%%%
%%%%%%%%%%%%%%%%%%%%%%%%%%%%%%%%%%%%%%%%%%%%%%%%%%%%%%%%%%%%%%%%%%%%%%%%%%%%

The mass of the fast quark plays only a minor role for its energy loss.
The right panel in Fig.~\ref{ratioLHC} shows (for LHC conditions, but 
similar statements apply at lower medium temperature) the asymptotic 
energy loss ratio for very high energy quarks as a function of the
quark mass. While the dynamical enhancement is largest for light
quarks, the difference between light and bottom quarks is only about 
$15\%$, and $b$ quarks still suffer about 70\% more energy loss in
a dynamical medium than in one with static scattering centers.

\begin{figure}[ht]
\vspace*{6cm} 
\includegraphics{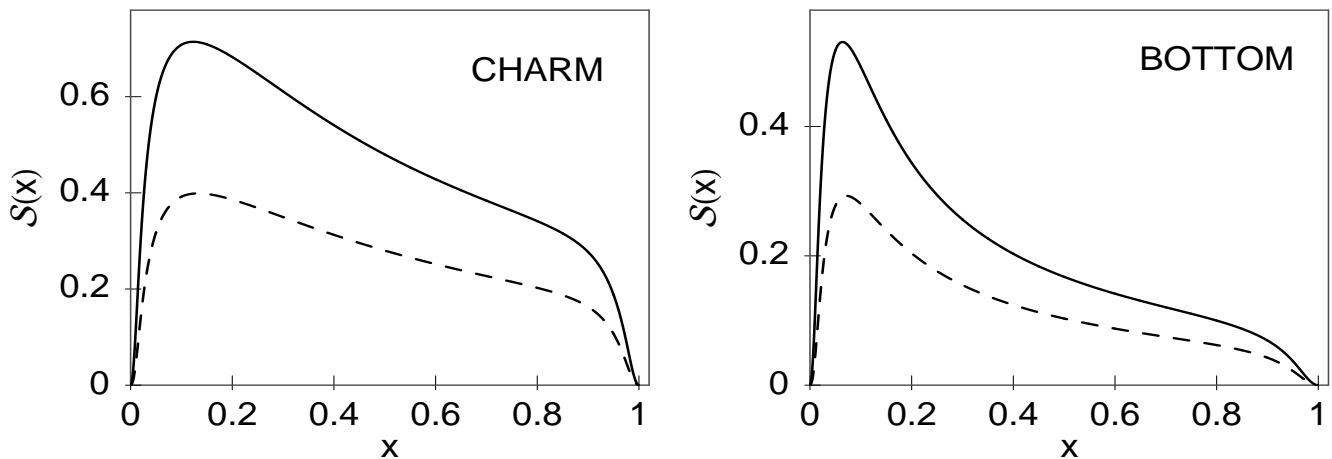}
\caption{Differential fractional radiative energy loss $\mathcal{S}(x)$ 
as a function of radiated energy fraction $x=\omega/E$ to first order in 
opacity, for charm (left) and bottom quarks (right) of initial energy 
$E=20$\,GeV traveling for $L=5$\,fm through a dynamical (solid lines) or 
static medium (dashed lines) of temperature $T=225$\,MeV.}
\label{xdep}
\end{figure}

\begin{figure}[ht]
\vspace*{9.5cm} 
\includegraphics{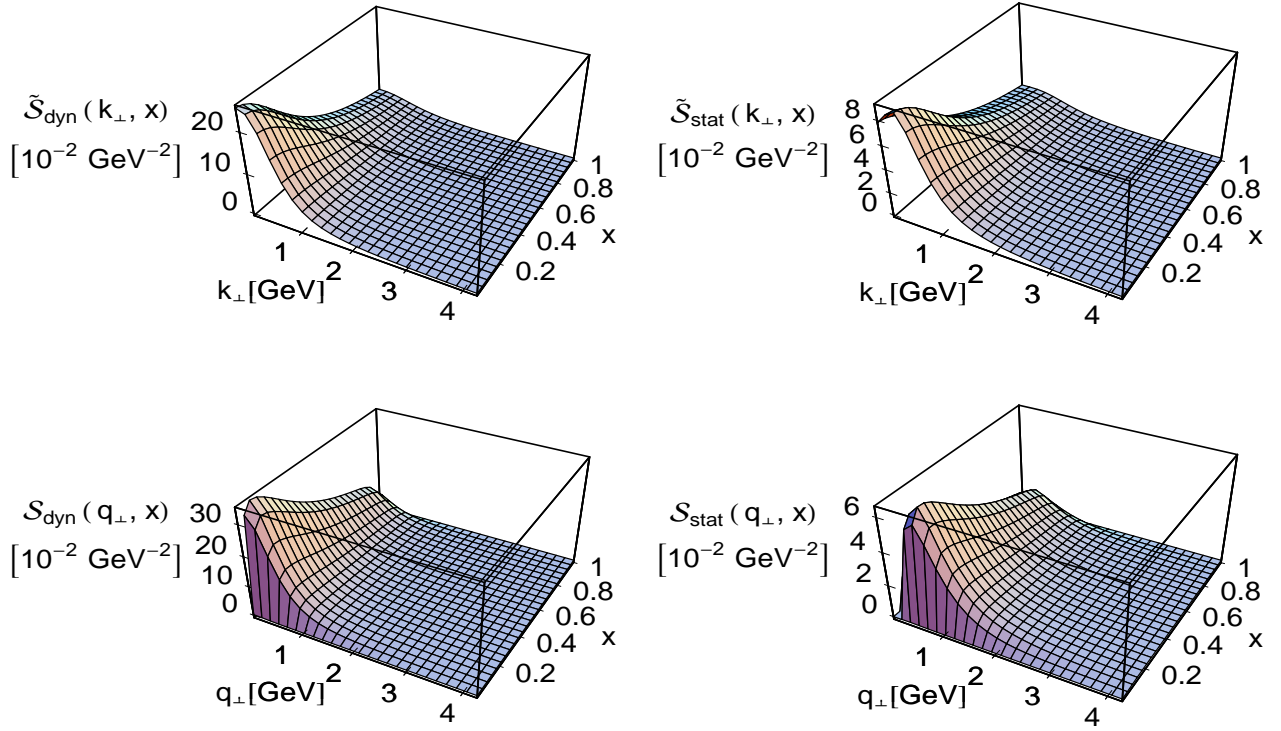}
\caption{(Color online) Charm quark radiative energy loss distributions 
$\tilde{\mathcal{S}}(\bk,x)$ (top panels; these are proportional to the 
emitted gluon spectra) and $\mathcal{S}(\bq,x)$ (bottom panels) for a 
dynamical (left) and a static medium (right). $x$ denotes the energy 
fraction $x=\omega/E$, and q$_\perp$ and k$_\perp$ stand for the transverse
momenta $|\bq|$ and $|\bk|$, respectively. The initial quark energy is
20\,GeV, the temperature of the medium $T=225$\,MeV, and a path length 
$L=5$~fm was assumed.}
\label{3Dspectrum_qk_Charm}
\end{figure}

\begin{figure}[htb]
\vspace*{9.5cm} 
\includegraphics{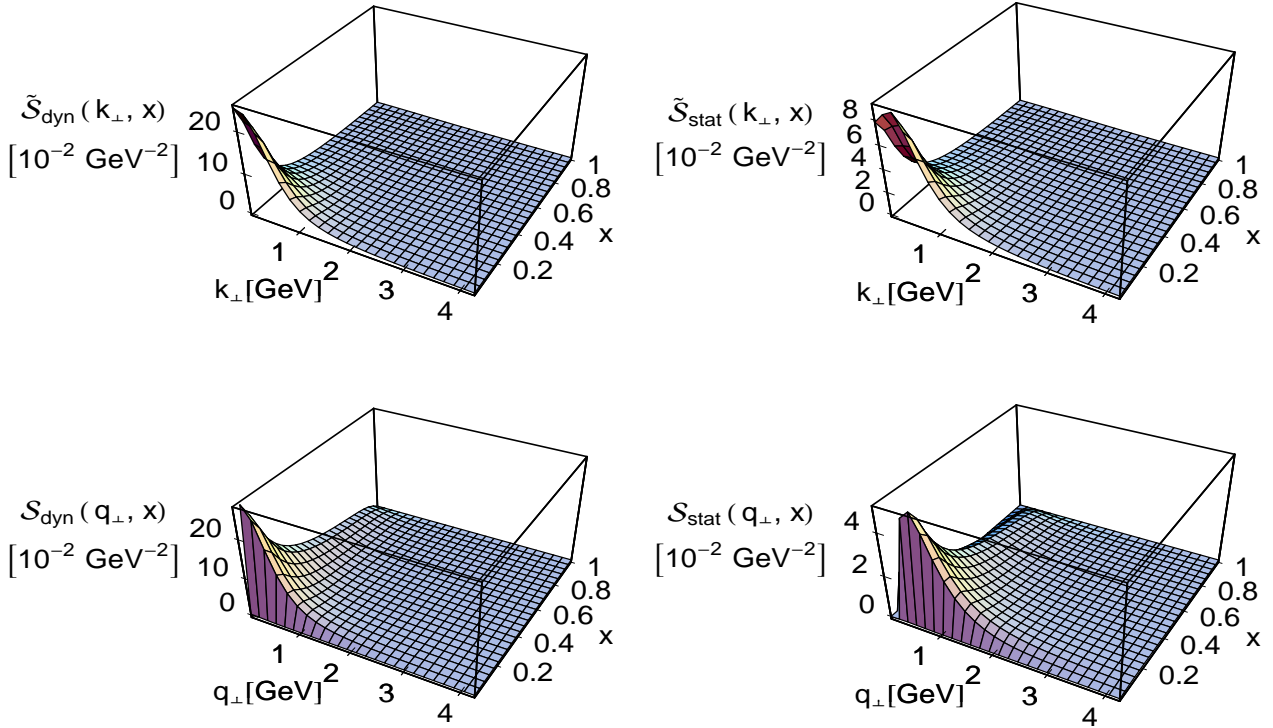}
\caption{(Color online) Same as Figure~\ref{3Dspectrum_qk_Charm}, but
for bottom quarks.}
\label{3Dspectrum_qk_Bottom}
\end{figure}

Stronger quark mass effects are seen in the shapes of the exchanged and
emitted gluon spectra themselves. Figure~\ref{xdep} shows the fractional
energy loss as a function of the energy fraction $x=\omega/E$ of the
radiated gluon relative to the initial quark energy. For both static
and dynamic media the fractional energy loss for bottom quarks (i.e.
the emitted gluon energy spectrum) is seen to be significantly softer 
%
%%%%%%%%%%%%%%%%%%%%%%%% Fig. 6 %%%%%%%%%%%%%%%%%%%%%%%%%%%%%%%%%%%%%%%%%%%%%
%%%%%%%%%%%%%%%%%%%%%%%%%%%%%%%%%%%%%%%%%%%%%%%%%%%%%%%%%%%%%%%%%%%%%%%%%%%%%
%
for bottom than for charm quarks. No such strong mass effect is visible in 
the shapes of the transverse momentum spectra of exchanged and emitted 
gluons: Figs.~\ref{3Dspectrum_qk_Charm} and~\ref{3Dspectrum_qk_Bottom} show
that, at fixed $x$, the exchanged and emitted gluon transverse momentum
%
%%%%%%%%%%%%%%%%%%%%%%%% Fig. 7 %%%%%%%%%%%%%%%%%%%%%%%%%%%%%%%%%%%%%%%%%%%%%
%%%%%%%%%%%%%%%%%%%%%%%%%%%%%%%%%%%%%%%%%%%%%%%%%%%%%%%%%%%%%%%%%%%%%%%%%%%%%
%
spectra have very similar shapes for charm and bottom quarks, and that the
main difference shows up in the $x$-dependence of these spectra. This 
strong quark mass effect on the emitted gluon energy spectrum is a 
consequence of the well known ``dead-cone effect''
\cite{Dokshitzer:2001zm,DG_Ind}. Except for very low $x$-values, the 
differential energy loss $\mathcal{S}(x)$ is a decreasing function of 
$\chi=M^2 x^2 +m_g^2$, and $\chi$ grows significantly faster with $x$
for bottom than for charm quarks. Thus, as $x$ increases, the contribution 
to the energy loss decreases more rapidly for bottom than for charm quarks. 
This is borne out by Fig.~\ref{xdep}.
%
%%%%%%%%%%%%%%%%%%%%%%%% Fig. 8 %%%%%%%%%%%%%%%%%%%%%%%%%%%%%%%%%%%%%%%%%%%%
%%%%%%%%%%%%%%%%%%%%%%%%%%%%%%%%%%%%%%%%%%%%%%%%%%%%%%%%%%%%%%%%%%%%%%%%%%%%%
%

In the present study, large-$x$ contributions to the total energy loss 
are not strongly suppressed, and even for bottom quarks the contribution 
from $x$ regions where the soft gluon approximation $\omega \ll E$ becomes 
doubtful could be as large as 30\%. This is mainly a deficiency of our 
approximations -- for a medium of static scatterers it is known 
that the LPM effect strongly suppresses the emission of large-$x$ gluons 
(see Eq.~(11) in the second paper of Ref.~\cite{DG_Ind}). Including
such effects should improve the applicability of our approximations,
by reducing large-$x$ gluon emission. That is, the total 
radiative energy loss should be reduced, without qualitatively affecting the 
energy loss ratio between static and dynamical media (since this ratio 
is seen in Fig.~\ref{xdep} to be largely independent of $x$).

%%%%%%%%%%%%%%%%%%%%%%%%%%%%%%%%%%%%%%%%%%%%%%%%%%%%%%%%%%%%%%%%%%%%%%%%
\section{Conclusion}
\label{sec5}
%%%%%%%%%%%%%%%%%%%%%%%%%%%%%%%%%%%%%%%%%%%%%%%%%%%%%%%%%%%%%%%%%%%%%%%%

Static scattering center approximation was used in all previous 
calculations of heavy quark radiative energy loss. An important consequence 
of this approximation is that it results in exactly zero collisional energy 
loss. However, it was recently found~\cite{Mustafa,Dutt-Mazumder,MD_Coll,Adil} 
that, under RHIC conditions, heavy quark collisional energy loss is significant 
and comparable to the previously calculated radiative energy loss. Since the 
static approximation is evidently inadequate in the computation of collisional 
energy loss, there arises a question whether such approximation is appropriate 
in the radiative energy loss case.

We here revisited the problem of heavy quark radiative energy loss, but now 
in dynamical medium of thermally distributed massless quarks and gluons. Our 
work has two goals: 1) To address the applicability of static approximation 
in radiative energy loss computations, and 2) To compute collisional and 
radiative energy losses within a consistent theoretical framework. 
In this paper we report the first step in this direction, 
where we compute the $1^{st}$ order in opacity contribution to the radiative 
energy loss in a dynamical QCD medium of infinite size.

We have shown that each individual contribution in the diagrammatic 
expansion of the energy loss in a dynamical medium is infrared divergent, 
due to the absence of magnetic screening \cite{Le_Bellac}. However, it is 
interesting that the sum of these contributions lead to an infrared safe 
result. The magnetic infrared divergence is thus naturally regulated, 
eliminating the need for introducing an artificial magnetic gluon mass 
when computing the radiative energy loss in a dynamical QCD medium.

The analytic expression for the radiative energy loss in a dynamical 
QCD medium was found to be remarkably similar to the one obtained in 
the static approximation. Still, the seemingly small differences, observed 
in the analytical expressions, were found to have important quantitative 
consequences: At the same (first) order in opacity, fast quarks that
propagate through a dynamical QCD medium lose energy at almost twice 
the rate computed for a medium of static scatterers. Recoil of the 
(massless) quarks and gluons in the medium is thus a phenomenologically 
important effect, which can not be neglected. We found no corners of the 
kinematic phase-space where the static scattering approximation is 
valid, neither for light nor for heavy fast quarks. Hence, the constituents 
of QCD medium can not be approximated as static scattering centers, and 
for reliable predictions of radiative energy loss, dynamical effects have 
to be included.

High precision heavy flavor measurements are expected to emerge from the 
upcoming high luminosity RHIC and LHC experiments. An important goal of 
heavy flavor energy loss measurements is to provide a tomographic 
diagnostic tool for the hot QCD matter created in these collisions. 
Therefore, reliable quantitative predictions for these experiments are 
essential. The results presented in this paper lead to the important 
qualitative conclusion that the observed quark energy loss could be 
significantly larger than previously thought. Turning this qualitative 
insight into a quantitative comparison with existing data, and reliable 
predictions for upcoming data requires, however, a significant additional
work. Most importantly, the present study does not take into account
coherent interference (LPM) effects and their modification by the
finite size of the medium created in heavy-ion collision fireballs.
The computation of heavy quark energy loss in a finite size dynamical 
QCD medium is therefore our next important goal.

\begin{acknowledgments}
The authors acknowledge the hospitality of the Institute of Nuclear Theory 
in Seattle where part of this work was done. Valuable discussions with Eric 
Braaten, Miklos Gyulassy, and Yuri Kovchegov are gratefully acknowledged. This 
work is supported by the U.S. Department of Energy, grant DE-FG02-01ER41190.
\end{acknowledgments}

\appendix

%%%%%%%%%%%%%%%%%%%%%%%%%%%%%%%%%%%%%%%%%%%%%%%%%%%%%%%%%%%%%%%%%%%%%%%%%%%%%
\section{Assumptions and approximations}
\label{appa}
%%%%%%%%%%%%%%%%%%%%%%%%%%%%%%%%%%%%%%%%%%%%%%%%%%%%%%%%%%%%%%%%%%%%%%%%%%%%%
\subsection{Kinematics}
\label{appa1}
%%%%%%%%%%%%%%%%%%%%%%%%%%%%%%%%%%%%%%%%%%%%%%%%%%%%%%%%%%%%%%%%%%%%%%%%%%%%%

In this paper we consider a heavy quark of mass $M$ which is produced
in the remote past on its mass shell, with large spatial momentum $p'\gg M$. 
We choose coordinates such that the momentum of the initial quark is 
along the $z$ axis:
\beqar
p'=\left(E'\approx p'+\frac{M^2}{2p'},\, p',\, \bm{0}\right) \; ,
\eeqar{pprime0}
We are interested in the radiative energy loss to first order in the 
opacity, so we study the case in which the quark exchanges (in arbitrary 
sequence) one virtual gluon with space-like momentum 
\beqar
  q = (q_0,\vq) = (q_0,\,q_z,\,\bq), \qquad q_0 \le |\vq|
\eeqar{q} 
with a parton in the medium and radiates one (medium-modified) real gluon 
with time-like momentum 
\beq
  k = ( k_0,\vk) = (\omega,k_z,\bk), \qquad k_0 \geq|\vk|
\eeq{k} 
into the medium. The quark emerges with 4-momentum $p^\mu$. 

\medskip

For the computation of the Feynmann diagrams given in Appendices B-D we 
will need cut propagators for the heavy quark $p$ ($D^> (p)$), the 
radiated gluon $k$ ($D^>_{\mu\nu} (k)$), and the exchanged gluon 
$q$ ($D^>_{\mu\nu} (q)$). 

The effective 1-HTL gluon propagators for the exchanged and emitted 
gluons have the form given in Eq.~(\ref{dmnMed}). By following the 
procedure outlined in \cite{Le_Bellac}, we obtain for the cut full 
gluon propagator
\beqar
D^>_{\mu\nu}(l)= - (1{+}f(l_0)) \Bigl( P_{\mu \nu} (l) \rho_T (l) +
Q_{\mu \nu} (l) \rho_L (l) \Bigr) \; ,
\eeqar{D>}
where $l$ is gluon momentum, $f(l_0)=(e^{l_0/T}{-}1)^{-1}$, and $T$ is the 
temperature of the medium. $\rho_{L,T}(l)$ are spectral functions defined 
by
%\beqar
%\rho_{L,T} (l) = 2 \pi \; \delta(l^2- \Pi_{T,L}(l)) - 
%\frac{2 {\rm Im} \Pi_{T,L}(l) \; \theta(1-\frac{l_0^2}{\vec{l}^2})}
%{(l^2{-}{\rm Re}\,\Pi_{T,L}(l))^2 + ({\rm Im}\,\Pi_{T,L}(l))^2} \;.
%\eeqar{rho_TL}
\beqar
\rho_{L,T} (l) = 2 \pi \; \delta(l^2- \Pi_{T,L}(l)) - 2 \, {\rm Im} 
\left (\frac{1}{l^2{-}\Pi_{T,L}(l)} \right )
\theta(1-\frac{l_0^2}{\vec{\bf l}^2}) \; .
\eeqar{rho_TL}
It was shown in \cite{DG_TM} that for the radiated gluon with momentum $k$
the longitudinal contribution can be neglected relative to the transverse 
one, and that for the transverse gluon the self energy $\Pi_T(k)$ can be 
approximated by $m_g^2$, where 
$m_g \approx \mu/\sqrt{2}=gT\sqrt{(6{+}n_f)/12}$ is the asymptotic mass. 
These approximations are true in the soft rescattering limit 
$\omega \gg |\bq|{\,\sim\,}|\bk|{\,\sim\,}g T$ which we use in this 
paper. With these approximations the HTL gluon propagator for the emitted 
gluon can be simply approximated by \cite{DG_TM}
\beqar
D_{\mu \nu}(k) \approx -i\, \frac{P_{\mu \nu}(k)}{k^2-m_g^2+i\epsilon}\; ,
\eeqar{rad_contrib}
where $P_{\mu \nu}$ is the transverse projector. The cut propagator for the 
radiated gluon is then given by~\cite{Le_Bellac,DG_TM}
\beqar
D_{\mu \nu}^>(k) \approx - \, 2 \pi (1+f(\omega)) \,
\frac{P_{\mu \nu}(k)}{2 \omega} \, \delta (k_0 - \omega) \; ,
\eeqar{rad_cut1}
where $\omega \approx \sqrt{\vk^2+m_g^2}$.

By using Eqs.~(\ref{kp},~\ref{x}) defined below, we obtain 
$f(\omega)=(e^{x E/T}{-}1)^{-1} \ll 1$ for highly energetic jets and 
$x>T/E$. Eq.~(\ref{rad_cut1}) can then be simplified to 
\beqar
D_{\mu \nu}^>(k) \approx - 2 \pi \, \frac{P_{\mu \nu}(k)}{2 \omega} \, \delta (k_0 - \omega) \;  .
\eeqar{rad_cut}
Similarly, the cut propagator for the heavy quark 
(with $D(p) = \frac{i}{p^2-M^2+i\epsilon}$) is given by
\beqar
D^> (p)= 2 \pi \frac{1}{2 E} \delta (p_0-E). 
\eeqar{quark_cut}

Unfortunately, the above approximations cannot be used for the 
virtual gluon mediating the collisional interaction. Both transverse and 
longitudinal contributions have to be kept in the gluon propagator $D(q)$, 
and it can be shown numerically that both contributions are equally 
important. Furthermore, no further simplifications can be made in the 
expressions for the transverse and longitudinal self energies $\Pi_T(q)$ 
and $\Pi_L(q)$ (see Eq.~(\ref{PiT})), beyond the restriction (\ref{q}) 
to space-like momenta. However, due to this restriction, the $\delta$ 
function in Eq.~(\ref{rho_TL}) does not contribute to the exchanged gluon 
spectral function, leading to
\beqar
D^>_{\mu\nu} (q)=\, \theta(1-\frac{q_0^2}{\vq^2})\, (1+f(q_0))\;
2 \, {\rm Im} \left( 
\frac{P_{\mu \nu} (q)}{q^2{-}\Pi_{T}(q)} + 
\frac{Q_{\mu \nu} (q)}{q^2{-}\Pi_{L}(q)}  \right)\; .
\eeqar{exchanged_cut}

%\beqar
%D^>_{\mu\nu} (q)=\, \theta(1-\frac{l_0^2}{\vec{l}^2})\, (1+f(q_0))\, 
%\Bigl( P_{\mu \nu} (q) \frac{2 {\rm Im} \Pi_{T}(l) }
%{(l^2{-}{\rm Re}\,\Pi_{T}(l))^2 + ({\rm Im}\,\Pi_{T}(l))^2} + 
%Q_{\mu \nu} (q) \frac{2 {\rm Im} \Pi_{L}(l)}
%{(l^2{-}{\rm Re}\,\Pi_{L}(l))^2 + ({\rm Im}\,\Pi_{L}(l))^2}\Bigr) \; .
%\eeqar{exchanged_cut}

As in \cite{Gyulassy_Wang,GLV,Wiedemann,WW,DG_Ind,ASW,MD_TR}, we assume validity
of the soft gluon ($\omega{\,\ll\,}E$) and soft rescattering 
($|\bq|{\,\sim\,}|\bk|{\,\ll\,}k_z$) approximations. Together with 
conservation of energy and momentum ($p'=p+k+q$) they yield
\beqar
k = \Bigl(\omega \approx k_z+\frac{\bk^2{+}m_g^2}{2 k_z},\,k_z,\,\bk\Bigr) ,
\qquad
p = \Bigl(E\approx p_z+\frac{(\bk{+}\bq)^2+M^2}{2 p_z},\,p_z,\,-(\bk{+}\bq)
\Bigr)
\; ,
\eeqar{kp}
and
\beqar
p' = \Bigl(E'\approx p_z{+}k_z{+}q_z{+}\frac{M^2}{2(p_z{+}k_z{+}q_z)},\,
p_z{+}k_z{+}q_z,\bm{0}\Bigr)  .
\eeqar{pprime}

In the next subsection we will show that it is reasonable to assume that 
$q_z$ has the same order of magnitude as $|\bq|$. Since $|\bk| \ll k_z$ 
and $q_z{\,\sim\,}|\bq|{\,\sim\,}|\bk|$, we then also have
$q_z{\,\ll\,}k_z$. Thus $k_z{+}q_z \approx k_z$ and
$p_z{+}k_z{+}q_z \approx p_z{+}k_z \approx p_z{+}q_z \approx p_z$.
Defining 
\beqar
  x \equiv \frac{k_z}{p_z} \; .
\eeqar{x}
we can further rewrite
\beq
 (p+k)^2-M^2 \approx \frac{\bk^2+M^2 x^2+m_g^2}{x} \approx M^2-(p'-k)^2
\eeq{ppprimek}
and show that
\beqar
  p^\mu P_{\mu\nu}(k) p^\nu \approx {p'}^\mu P_{\mu\nu}(k) p^\nu 
= p^\mu P_{\mu\nu}(k) {p'}^\nu \approx {p'}^\mu P_{\mu\nu}(k) {p'}^\nu
  \approx - \frac{\bk^2}{x^2}\, ,
\eeqar{pPp}
where $P_{\mu\nu}(k)$ is a transverse projector of radiated gluon, defined by Eq.~(\ref{PQmunu}).

Finally, by using Eqs.~(\ref{q}) and (\ref{kp})--(\ref{pprime}), we obtain
\beqar
  E'-E-\omega-q_0\approx q_z-q_0-\frac{\bk^2+M^2 x^2+m_g^2}{2xE} 
  \approx q_z-q_0\,.
\eeqar{Energy_difference}

%%%%%%%%%%%%%%%%%%%%%%%%%%%%%%%%%%%%%%%%%%%%%%%%%%%%%%%%%%%%%%%%%%%%%%%%%%%%%
\subsection{$\bm{q_z}$ vs. $|\bq|$ comparison}
\label{appa2}
%%%%%%%%%%%%%%%%%%%%%%%%%%%%%%%%%%%%%%%%%%%%%%%%%%%%%%%%%%%%%%%%%%%%%%%%%%%%%

\begin{figure}[ht]
\vspace*{5.8cm} 
\includegraphics{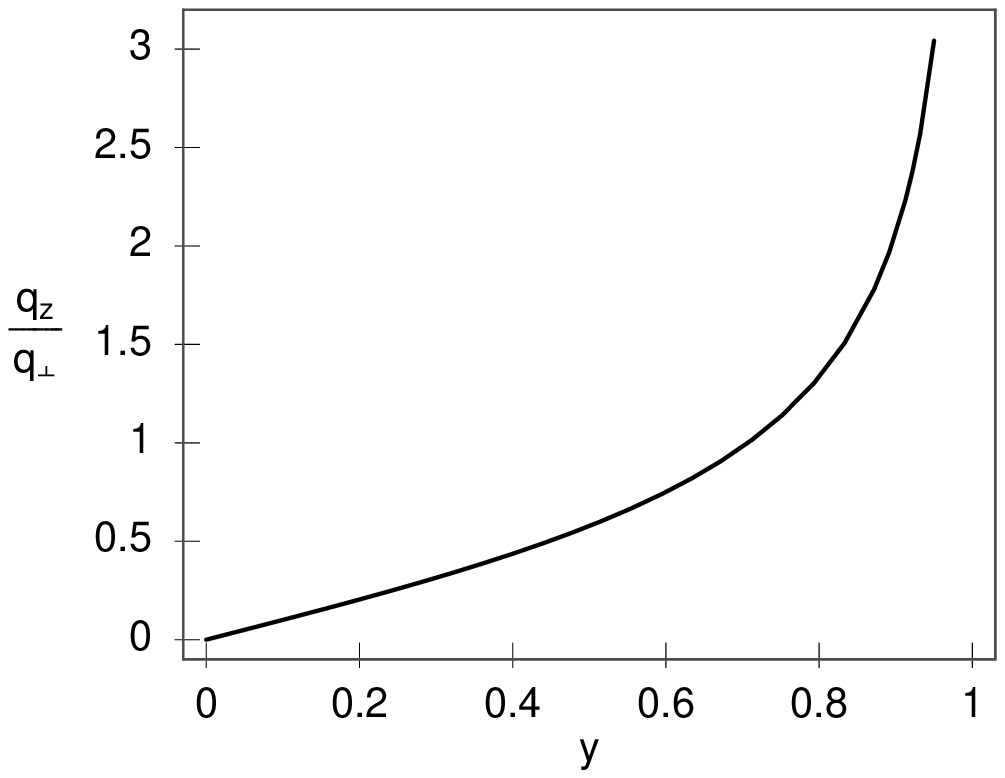}
\includegraphics{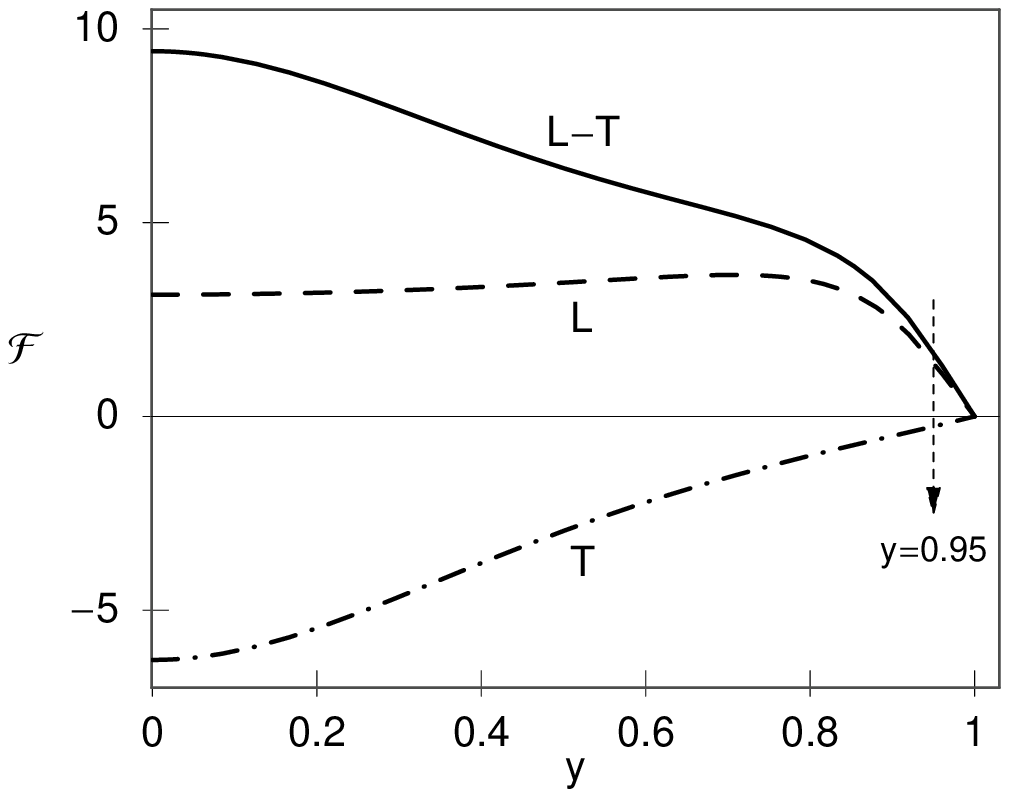}
\caption{Left: The ratio $\frac{q_z}{q_\perp}\equiv\frac{q_z}{|\bq|}$ 
as a function of $y$. Right: Transverse (dot-dashed) and longitudinal 
(dashed) contributions to the energy loss from virtual gluons with
typical transverse momenta $|\bq|=0.5$\,GeV, as functions of $y$. Full 
curve shows the difference between longitudinal and transverse contributions 
(see Eq.~(\ref{I_q1})). The vertical arrow indicates the $y$-region above 
which $q_z$ exceeds $|\bq|$ by more than a factor 3.}
\label{qzqt_vs_y}
\end{figure}

Equation (\ref{Energy_difference}) together with energy conservation 
implies $q_0 \sim q_z$. Introducing the variable  
$y=\frac{q_0}{\sqrt{q_z^2+\bq^2}}$ (with $-1\le y \le 1$), we can 
further express $q_z$ in terms of $\bq^2$:
\beqar
  q^2_z= \bq^2 \frac{y^2}{1-y^2}\,.
\eeqar{qz_vs_qperp}
The left panel in Fig.~\ref{qzqt_vs_y} shows the ratio $q_z/q_\perp
= q_z/\sqrt{\bq^2}$ over the entire $y$ range. We see that, except
for the region $y\to1$, $q_z$ and $|\bq|$ are comparable, and that
for $|y| < 0.95$ the ratio $\frac{q_z}{|\bq|}$ remains below 3. Hence,
for $|y| < 0.95$, $q_z$ and $|\bq|$ have the same order of magnitude.

%%%%%%%%%%%%%%%%%%%%%%%%%%% Fig. 9 %%%%%%%%%%%%%%%%%%%%%%%%%%%%%%%%%%%%%%%%%
%%%%%%%%%%%%%%%%%%%%%%%%%%%%%%%%%%%%%%%%%%%%%%%%%%%%%%%%%%%%%%%%%%%%%%%%%%%%

We next want to test how important the region $|y| > 0.95$ is for the energy loss. To do this, we
start from the following equation:
\beqar
\mathcal{F}_{T,L}(y)=\frac{1}{2\pi}\, \frac{1}{y}\, 
\frac{2\,{\rm Im}\,\Pi_{T,L}(y)}
     {\bigl(\bq^2+{\rm Re}\,\Pi_{T,L}(y)\bigr)^2 
    + \bigl({\rm Im}\,\Pi_{T,L}(y)\bigr)^2} \; ,
\eeqar{F} 
which gives $y$-integrand of the transverse and longitudinal contributions 
to the energy loss (see Eqs.~(\ref{I_q1}) and (\ref{Sum_rule})). By using 
Eq.~(\ref{F}) we obtain $F_{T,L} (y)$ for a typical transverse 
momentum $|\bq|=0.5$\,GeV of the exchanged gluon, which is shown 
in the right panel of Fig.~\ref{qzqt_vs_y}.

We see that the main contribution to the energy loss comes from the region 
$|y| < 0.95$, especially after accounting that the contribution to the energy loss 
comes from the difference between longitudinal and transverse integrands 
(see Eq.~(\ref{I_q1}) and the full curve in the right panel of Fig.~\ref{qzqt_vs_y}). 
We additionally tested that the error made by computing the energy loss using 
our approximation in the region $|y| > 0.95$ (where the approximation breaks 
down) is less than~2\%.

%%%%%%%%%%%%%%%%%%%%%%%%%%%%%%%%%%%%%%%%%%%%%%%%%%%%%%%%%%%%%%%%%%%%%%%%%%%%%
\section{Computation of diagrams $\bm{M_{1,0,1}}$ - $\bm{M_{1,0,4}}$}
\label{appb}
%%%%%%%%%%%%%%%%%%%%%%%%%%%%%%%%%%%%%%%%%%%%%%%%%%%%%%%%%%%%%%%%%%%%%%%%%%%%%

\begin{figure}[ht]
\vspace*{9.cm} 
\includegraphics{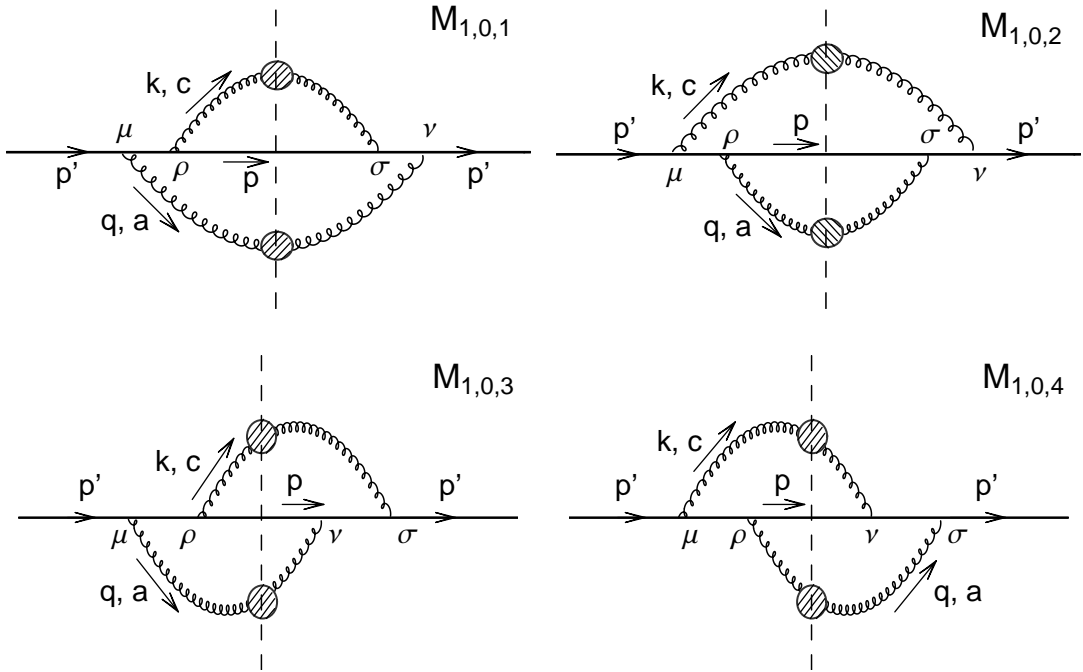}
\caption{Feynman diagrams $M_{1,0,1}$, $M_{1,0,2}$, $M_{1,0,3}$ and 
$M_{1,0,4}$ contributing to the radiative energy loss to first order in 
opacity. The large dashed circles (``blob'') represent effective HTL
gluon propagators \cite{DG_TM}. A cut gluon propagator with momentum 
$k$ and color $c$ corresponds to the radiated gluon ($\omega>|\vk|$). 
A cut gluon propagator with momentum $q$ and color $a$ corresponds to 
a collisional interaction with a parton in the medium ($q_0 \le |\vq|$).}
\label{DiagM10}
\end{figure}

In this appendix we present in some detail the calculation of the diagrams 
shown in Fig.~\ref{DiagM10}. These diagrams present contributions where
both ends of the exchanged gluon $q$ are attached to the heavy quark,
i.e. none is attached to the radiated gluon $k$ and no 3-gluon vertex 
is involved. 

Here and later the diagrams are labeled as follows: In $M_{1,i,j}$, 
$1$ denotes that these diagrams contribute to the energy loss to first 
order in opacity; $i$ denotes how many ends of the virtual gluon $q$ 
are attached to the radiated gluon $k$; and $j$ labels the specific
diagram in that class.

%%%%%%%%%%%%%%%%%%%%%%%% Fig. 10 %%%%%%%%%%%%%%%%%%%%%%%%%%%%%%%%%%%%%%%%%%%%
%%%%%%%%%%%%%%%%%%%%%%%%%%%%%%%%%%%%%%%%%%%%%%%%%%%%%%%%%%%%%%%%%%%%%%%%%%%%%

{\bf 1.} We will first calculate the cut diagram 
$M_{1,0,1}^>=2 \, {\rm Im}\,M_{1,0,1}$~\cite{Le_Bellac}:
\beqar
M_{1,0,1}^> &=& \int \, (-ig (2p'{-}q)^\mu) D^>_{\mu\nu}(q) 
(ig (2p'{-}q)^\nu) \, \frac{i}{(p{+}k)^2-M^2 + i \epsilon} \,
\frac{-i}{(p{+}k)^2-M^2 - i \epsilon} \,
(-ig (2p{+}k)^\rho) D^>_{\rho\sigma}(k) (ig (2p{+}k)^\sigma) 
\nonumber \\
&& \; \; \times\  D^> (p)\; t_a t_c t_c t_a \; (2 \pi)^4 
\delta^{(4)}(p'{-}p{-}k{-}q)\;
\frac{d^4 p}{(2\pi)^4} \frac{d^4q}{(2\pi)^4} \frac{d^4k}{(2\pi)^4} \;,
\eeqar{M101_1} 
where $D^>_{\rho\sigma}(k)$, $D^> (p)$ and $D^>_{\mu\nu}(q)$ (given by 
Eqs.~(\ref{rad_cut}),~(\ref{quark_cut}) and~(\ref{exchanged_cut}))
are the cut propagators for the radiated gluon, the heavy quark, and 
the exchanged gluon, respectively. $M_{1,0,1}^>$ then becomes

\beqar
M_{1,0,1}^> &=& \int g^4 \frac{1}{(p{+}k)^2-M^2 + i \epsilon} \,
\frac{1}{(p{+}k)^2-M^2 - i \epsilon} \,
(2p{+}k)^\rho P_{\rho\sigma}(k) (2p{+}k)^\sigma (- 2 \pi) \, 
\frac{\delta (k_0 - \omega)}{2 \omega} \, \nonumber\\
&& \; \; \times \; \theta(1-\frac{q_0^2}{\vq^2})\, (1+ f(q_0)) \,
(2p'{-}q)^\mu \, 2 \, {\rm Im} \left( 
\frac{P_{\mu \nu} (q)}{q^2{-}\Pi_{T}(q)} + 
\frac{Q_{\mu \nu} (q)}{q^2{-}\Pi_{L}(q)}  \right) (2p'{-}q)^\nu
\nonumber \\
&&\quad\times\  
t_a t_c t_c t_a \,2 \pi \frac{\delta (p_0-E)}{2 E}\, 
2\pi\delta(p'_0{-}p_0{-}k_0{-}q_0)\,
\frac{dp_0}{2\pi}\frac{d^4q}{(2\pi)^4} \frac{d^4 k}{(2 \pi)^4} \;,
\eeqar{M101_2} 
where we have used three of the $\delta$-functions
to do the integral over $d^3p$. Correspondingly, it should be kept in mind
that in (\ref{M101_2}) spatial components of $p$ should be replaced by the
corresponding components of $p'{-}k{-}q$ which can be then further simplified
with the approximations discussed in Appendix \ref{appa1}. The same will be
understood when evaluating the other three diagrams further below. 

From Eq.~(\ref{pPp}) and the fact that $k^\rho P_{\rho\sigma}(k)=0$
we obtain 
\beqar
(2p{+}k)^\rho P_{\rho\sigma}(k) (2p{+}k)^\sigma
\approx - 4\, \frac{\bk^2}{x^2}\,. 
\eeqar{rad_contrib1}
For highly energetic jets, and by using Eqs.~(\ref{PQmunu}) 
and~(\ref{Energy_difference}), we obtain
$(2p'{-}q)^\mu P_{\mu\nu}(q) (2p'{-}q)^\nu
\approx - \, (2p'{-}q)^\mu Q_{\mu\nu}(q) (2p'{-}q)^\nu 
\approx - \, 4 E'^2 \bq^2/\vq^2$, which leads to  
\beqar
(2p'{-}q)^\mu \, 2 \, {\rm Im} \left( 
\frac{P_{\mu \nu} (q)}{q^2{-}\Pi_{T}(q)} + 
\frac{Q_{\mu \nu} (q)}{q^2{-}\Pi_{L}(q)}  \right) (2p'{-}q)^\nu
\approx  4 E'^2\, \frac{\bq^2}{\vq^2}\;  2 \, {\rm Im} 
\left( \frac{1}{q^2{-}\Pi_{L}(q)} - \frac{1}{q^2{-}\Pi_{T}(q)}   \right)
\eeqar{coll_contrib1}
By also using Eqs.~(\ref{rad_contrib1}) and (\ref{coll_contrib1}), and
after performing integrations over $p_0$ and $k_0$, Eq.~(\ref{M101_2}) 
reduces to
\beqar
M_{1,0,1}^> &=& g^4 \, t_a t_c t_c t_a \, \int \frac{1}{((p{+}k)^2-M^2)^2} \, 
 \frac{4 \bk^2}{x^2} \; \theta(1-\frac{q_0^2}{\vq^2})\, 
 (4 E'^2)\, \frac{\bq^2}{\vq^2}  \,(1+ f(q_0))\, 2 \, {\rm Im} 
\left( \frac{1}{q^2{-}\Pi_{L}(q)} - \frac{1}{q^2{-}\Pi_{T}(q)}   \right)
\nonumber \\
&&\quad\times\  
\frac{1}{2 E}  \, 2\pi\delta(p'_0{-}E{-}k_0{-}q_0)\,
\frac{d^4q}{(2\pi)^4} \frac{d^3 k}{(2 \pi)^3 2 \omega} \,.
\eeqar{M101_21}

We now use Eqs.~(\ref{ppprimek}), (\ref{Energy_difference}), as well as 
$\omega=\sqrt{\vec{\mathbf{k}}^2+m_g^2} \approx xE$ and $4E'^2/2E\approx 2E$,
to obtain:
\beqar
M_{1,0,1}^> &=&  8E\, g^4\, t_a t_c t_c t_a 
\int \frac{d^3k}{(2 \pi)^3 2 \omega}\, 
\frac{\bk^2}{(\bk^2{+}M^2 x^2{+}m_g^2)^2} 
\int \frac{d^4 q}{(2 \pi)^4}\, 2\pi\delta(q_0{-}q_z)\,  
\nonumber \\
&& \hspace*{0.3cm} \,\times\, (1+f(q_0))\, 
\frac{\bq^2}{\vq^2} \, 
\left(\frac{2\,{\rm Im}\,\Pi_L(q)}
           {(q^2{-}{\rm Re}\,\Pi_L(q))^2 + ({\rm Im}\,\Pi_L(q))^2}
    - \frac{2\,{\rm Im}\,\Pi_T(q)}
           {(q^2{-}{\rm Re}\,\Pi_T(q))^2 + ({\rm Im}\,\Pi_T(q))^2}
\right) ,
\eeqar{M101_4}   
where $f(q_0)=(e^{q_0/T}{-}1)^{-1}$ and $T$ is the temperature of the 
medium. For small $q_0$ we can expand
\beqar
  1+f(q_0) \approx \frac{T}{q_0}+\frac{1}{2} 
                 + {\cal O}\Bigl(\frac{q_0}{T}\Bigr).
\eeqar{f_app}
With this approximation Eq.~(\ref{M101_4}) becomes
\beqar
 M_{1,0,1}^> &=& 2E\,g^4\, t_a t_c t_c t_a 
  \int \frac{d^3k}{(2 \pi)^3 2\omega} \, 
  \frac{4\,\bk^2}{(\bk^2{+}M^2 x^2{+}m_g^2)^2}\, I_q
\eeqar{M101_5}
where $I_q$ is given by
\beqar
  I_q &=& \int \frac{d^4 q}{(2 \pi)^4}\, 2\pi\delta(q_0{-}q_z)\, 
  \frac{\bq^2}{\vq^2}\left(\frac{1}{2}+\frac{T}{q_0}\right) 
%\nonumber \\ && \hspace*{0.2cm} \; \times \; 
  \left(\frac{2\,{\rm Im}\,\Pi_L(q)}
             {(q^2{-}{\rm Re}\,\Pi_L(q))^2 + ({\rm Im}\,\Pi_L(q))^2}
      - \frac{2\,{\rm Im}\,\Pi_T(q)}
             {(q^2{-}{\rm Re}\,\Pi_T(q))^2 + ({\rm Im}\,\Pi_T(q))^2}
  \right)
\nonumber \\ 
       &=& \int \frac{dq_z\, d^2q}{(2 \pi)^3}\, \frac{\bq^2}{q_z^2{+}\bq^2}
  \left(\frac{1}{2}+\frac{T}{q_z}\right)
%\nonumber \\ && \hspace*{0.2cm} \; \times \; 
  \left(\frac{2\,{\rm Im}\,\Pi_L(q)}
             {(\bq^2{-}{\rm Re}\,\Pi_L(q))^2 + ({\rm Im}\,\Pi_L(q))^2}
      - \frac{2\,{\rm Im}\,\Pi_T(q)}
             {(\bq^2{-}{\rm Re}\,\Pi_T(q))^2 + ({\rm Im}\,\Pi_T(q))^2}
  \right) \; .
\eeqar{I_q}   
With the help of Eq.~(\ref{qz_vs_qperp}) and
\beqar
  dq_z &=& |\bq|\frac{dy}{(1{-}y^2)^{\frac{3}{2}}} \,,
\eeqar{q_z}
and noting that the polarization functions $\Pi_{T,L}$ depend only on 
$y$ and that $2\,{\rm Im}\,\Pi_{T,L}(y)/[(\bq^2{+}{\rm Re}\,\Pi_{T,L}(y))^2
+ ({\rm Im}\,\Pi_{T,L}(y))^2]$ is an odd function of this variable,
we can rewrite Eq.~(\ref{I_q}) as
\beqar
 I_q &=& T \int\frac{d^2q}{(2\pi)^2} \frac{1}{2\pi} \int_{-1}^{1} \frac{dy}{y} 
 \left(\frac{2\,{\rm Im}\,\Pi_L(y)}
            {(\bq^2{+}{\rm Re}\,\Pi_L(y))^2 + ({\rm Im}\,\Pi_L(y))^2}
     - \frac{2\,{\rm Im}\,\Pi_T(y)}
            {(\bq^2{+}{\rm Re}\,\Pi_T(y))^2 + ({\rm Im}\,\Pi_T(y))^2}
 \right) 
\nonumber \\
 &=& T \int \frac{d^2q}{(2\pi)^2}\, \frac{\mu^2}{\bq^2 (\bq^2{+}\mu^2)} \; .
\eeqar{I_q1}
Here we used the sum rules \cite{Aurenche}
\beqar
  \int_{-1}^{1} \frac{dy}{y}\, \frac{1}{2\pi}\, 
  \frac{2\,{\rm Im}\,\Pi_{T,L}(y)}
       {(\bq^2{+}{\rm Re}\,\Pi_{T,L}(y))^2 + ({\rm Im}\,\Pi_{T,L}(y))^2}
 =\left( \frac{1}{\bq^2+{\rm Re}\,\Pi_{T,L}(y{=}\infty)}
        -\frac{1}{\bq^2+{\rm Re}\,\Pi_{T,L}(y{=}0)}\right) 
\eeqar{Sum_rule}
with
\beqar
 {\rm Re}\,\Pi_{T,L}(y{=}\infty) = \frac{\mu^2}{3}\; , 
 \hspace*{0.5cm} 
 {\rm Re}\,\Pi_{T}(y{=}0)=0\; , 
 \hspace*{0.5cm} 
 {\rm Re}\,\Pi_{L}(y{=}0)=\mu^2\; . 
\eeqar{PTLy}
Finally, Eq.~(\ref{M101_5}) becomes ($2\,{\rm Im}\,M_{1,0,1} =M^>_{1,0,1}$)
\beqar
2\,{\rm Im}\,M_{1,0,1} = 8 E \, g^4 T  \, t_a t_c t_c t_a 
\int \frac{d^3k}{(2\pi)^3 2 \omega} \, \frac{d^2q}{(2\pi)^2} \, 
\frac{\bk^2}{(\bk^2{+}M^2 x^2{+}m_g^2)^2}\, 
\frac{\mu^2}{\bq^2 (\bq^2{+}\mu^2)} \,.
\eeqar{M101_final}

\bigskip

{\bf 2.} Next we consider the diagram $M^>_{1,0,2} = 2\,{\rm Im}\,M_{1,0,2}$:
\beqar
  M^>_{1,0,2} &=& \int \, (-i g (2p'{-}k)^{\mu}) \, D^>_{\mu \nu}(k) \, 
  (i g (2p'{-}k)^{\nu}) \, \frac{i}{(p'{-}k)^2-M^2 + i \epsilon} \,
  \frac{-i}{(p'{-}k)^2-M^2 - i \epsilon} \,  
\nonumber \\
  && \; \; \times (-i g (2p{+}q)^{\rho})\, D^>_{\rho \sigma} (q) \,(i g (2p{+}q)^{\sigma}) \,  D^> (p) \; t_c t_a t_a t_c \, 
     (2\pi)^4 \delta^{(4)}(p'{-}p{-}k{-}q)\;
     \frac{d^4p}{(2\pi)^4} \frac{d^4q}{(2\pi)^4} \frac{d^4k}{(2\pi)^4} \; .
\eeqar{M102_1}  
By applying the same techniques as above and using Eq.~(\ref{ppprimek}) 
we obtain
\beqar
  2\,{\rm Im}\,M_{1,0,2} = 8 E \, g^4 T \, t_c t_a t_a t_c 
  \int \frac{d^3k}{(2\pi)^3 2\omega} \, \frac{d^2q}{(2\pi)^2}\, 
  \frac{\bk^2}{(\bk^2{+}M^2 x^2{+}m_g^2)^2} \, 
  \frac{\mu^2}{\bq^2 (\bq^2{+}\mu^2)} \,.
\eeqar{M102_final}

\bigskip

{\bf 3.} Let us now compute the diagram $M^>_{1,0,3} = 2\,{\rm Im}\,M_{1,0,3}$:
\beqar
  M^>_{1,0,3} &=& \int (-ig (2p'{-}q)^{\mu})\, D^>_{\mu \nu}(q)\, 
  (ig (2p{+}q)^{\nu}) 
  \, \frac{i}{(p{+}k)^2-M^2 + i \epsilon} \, 
  (-ig (2p{+}k)^{\rho})\, D^>_{\rho \sigma}(k) \, 
  ( ig (2p'{-}k)^{\sigma}) \, 
\nonumber \\
  && \; \; \times \; \frac{-i}{(p'{-}k)^2-M^2 - i \epsilon} \, 
  t_a t_c t_a t_c \, D^> (p)\, 
 (2\pi)^4 \delta^{(4)}(p'{-}p{-}k{-}q)\,\frac{d^4p}{(2\pi)^4} 
 \frac{d^4q}{(2\pi)^4} \frac{d^4k}{(2\pi)^4} 
\nonumber \\
&=& g^4 \, t_a t_c t_a t_c \, \int \frac{1}{(p{+}k)^2-M^2 + i \epsilon} \,
\frac{1}{(p'{-}k)^2-M^2 - i \epsilon} \, 
(2p{+}k)^\rho P_{\rho\sigma}(k) (2p'{-}k)^\sigma (- 2 \pi) \, 
\frac{\delta (k_0 - \omega)}{2 \omega} \, \nonumber\\
&& \; \; \times \; \theta(1-\frac{q_0^2}{\vq^2})\, (1+ f(q_0)) \,
(2p'{-}q)^\mu \, 2 \, {\rm Im} \left( 
\frac{P_{\mu \nu} (q)}{q^2{-}\Pi_{T}(q)} + 
\frac{Q_{\mu \nu} (q)}{q^2{-}\Pi_{L}(q)}  \right) (2p{+}q)^\nu
\nonumber \\
&&\quad\times\  
2 \pi \frac{\delta (p_0-E)}{2 E}\, 2\pi\delta(p'_0{-}p_0{-}k_0{-}q_0)\,
\frac{dp_0}{2\pi}\frac{d^4q}{(2\pi)^4} \frac{d^4 k}{(2 \pi)^4} \;,
\eeqar{M103_1}
where we have used Eqs.~(\ref{rad_cut}), (\ref{quark_cut}) and 
(\ref{exchanged_cut}). We also used three of the $\delta$-functions
to do the integral over $d^3p$. By using 
Eqs.~(\ref{ppprimek})--(\ref{Energy_difference}), the cut amplitude of 
diagram $M_{1,0,3}$ becomes
\beqar
  2\,{\rm Im}\,M_{1,0,3} &=& 2E\, g^4\,t_a t_c t_a t_c 
  \int\frac{d^3k}{(2\pi)^3 2\omega}\, 
  \frac{4\bk^2}{(\bk^2{+}M^2 x^2{+}m_g^2)^2}\,\int\frac{d^4q}{(2\pi)^4}\,
  2\pi\delta(q_0{-}q_z)\,(1+f(q_0))\,\frac{\bq^2}{\vq^2} 
\nonumber \\
&& \hspace*{1cm} \; \times \; 
\left(\frac{2\,{\rm Im}\,\Pi_T(q)}
           {(q^2{-}{\rm Re}\,\Pi_T(q))^2 + ({\rm Im}\,\Pi_T(q))^2}
    - \frac{2\,{\rm Im}\,\Pi_L(q)}
           {(q^2{-}{\rm Re}\,\Pi_L(q))^2 + ({\rm Im}\,\Pi_L(q))^2}
\right)
\nonumber \\
&=& -8E \, g^4 \, t_a t_c t_a t_c \int \frac{d^3k}{(2\pi)^3 2\omega} 
    \,\frac{\,\bk^2}{(\bk^2{+}M^2 x^2{+}m_g^2)^2} \, I_q \,,
\eeqar{M103_2} \
where $I_q$ is given by Eq.~(\ref{I_q1}), giving finally 
\beqar
2\,{\rm Im}\,M_{1,0,3} = -8E\, g^4T\, t_a t_c t_a t_c  
\int \frac{d^3k}{(2\pi)^3 2\omega} \, \frac{d^2q}{(2\pi)^2} 
\,\frac{\,\bk^2}{(\bk^2{+}M^2 x^2{+}m_g^2)^2}\,
  \frac{\mu^2}{\bq^2 (\bq^2{+}\mu^2)} \,.
\eeqar{M103_final}

\bigskip

{\bf 4.} In the same way we obtain
\beqar
 2\,{\rm Im}\,M_{1,0,4} = -8E \, g^4T  \, t_c t_a t_c t_a 
 \int \frac{d^3k}{(2\pi)^3 2\omega} \, \frac{d^2q}{(2\pi)^2} 
 \,\frac{\,\bk^2} {(\bk^2{+}M^2 x^2{+}m_g^2)^2}\, 
   \frac{\mu^2}{\bq^2 (\bq^2{+}\mu^2)} \,.
\eeqar{M104_final}

{\bf 5.} The sum of all four diagrams (\ref{M101_final}), (\ref{M102_final}),
(\ref{M103_final}), and (\ref{M104_final}) thus becomes
\beqar
 2\,{\rm Im}\,M_{1,0} &\equiv& 2\,{\rm Im}\,M_{1,0,1} 
 + 2\,{\rm Im}\,M_{1,0,2} + 2\,{\rm Im}\,M_{1,0,3} + 2\,{\rm Im}\,M_{1,0,4} 
\nonumber \\
&=& 8E \, g^4T \, [t_a, t_c] \,[t_c, t_a] \int \frac{d^3k}{(2\pi)^3 2\omega}
\,\frac{d^2q}{(2\pi)^2}\, \frac{\bk^2}{(\bk^2{+}M^2 x^2{+}m_g^2)^2}\, 
\frac{\mu^2}{\bq^2 (\bq^2{+}\mu^2)}\;,
\eeqar{M10_final}
where $[t_a,t_c]$ is a color commutator.

%%%%%%%%%%%%%%%%%%%%%%%%%%%%%%%%%%%%%%%%%%%%%%%%%%%%%%%%%%%%%%%%%%%%%%%%%%%%%
\section{Computation of diagrams $\bm{M_{1,1,1}}$ - $\bm{M_{1,1,4}}$}
\label{appc}
%%%%%%%%%%%%%%%%%%%%%%%%%%%%%%%%%%%%%%%%%%%%%%%%%%%%%%%%%%%%%%%%%%%%%%%%%%%%%

\begin{figure}[ht]
\vspace*{8.5cm} 
\includegraphics{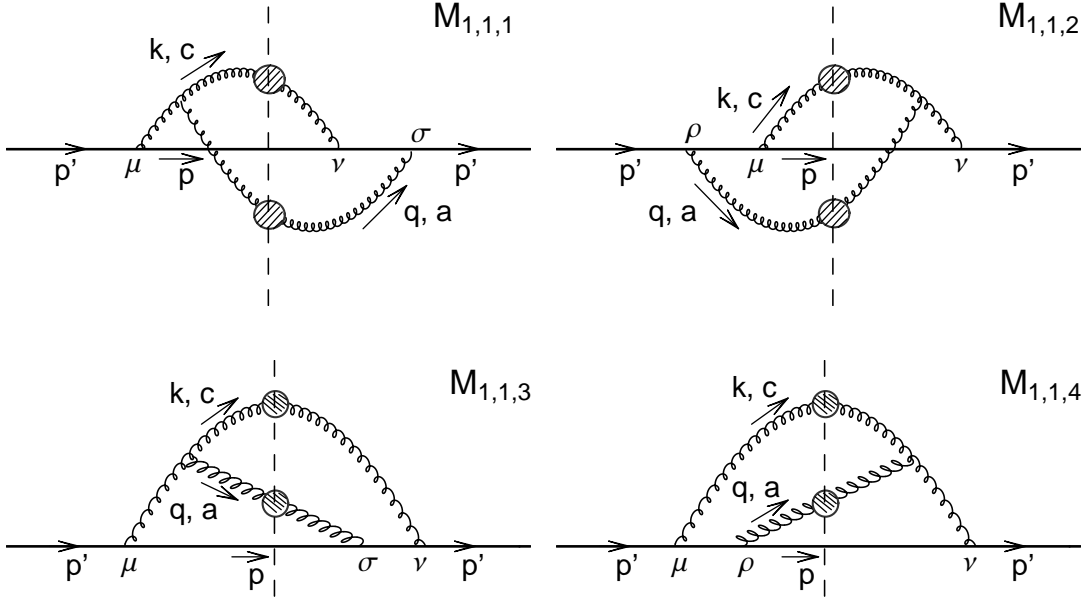}
\caption{Feynman diagrams $M_{1,1,1}$, $M_{1,1,2}$, $M_{1,1,3}$ and 
$M_{1,1,4}$ contributing to the radiative energy loss to first order in 
opacity, labeled in the same way as Fig.~\ref{DiagM10}.}
\label{DiagM11}
\end{figure}

In this Appendix we calculate the diagrams shown in Fig.~\ref{DiagM11}
where one of the ends of the exchanged gluon $q$ is attached to the 
radiated gluon $k$. 
%
%%%%%%%%%%%%%%%%%%%%%%%% Fig. 8 %%%%%%%%%%%%%%%%%%%%%%%%%%%%%%%%%%%%%%%%%%%%
%%%%%%%%%%%%%%%%%%%%%%%%%%%%%%%%%%%%%%%%%%%%%%%%%%%%%%%%%%%%%%%%%%%%%%%%%%%%
%

We start off with $M^>_{1,1,1}=2\,{\rm Im}\,M_{1,1,1}$:
\beqar
M^>_{1,1,1} &=& \int (-ig (2p'{-}k')^\mu t_b)\, D_{\mu\rho}(k')\, gf^{cba} 
\Bigl(g^{\rho\tau}(k'{+}q)^\lambda + g^{\lambda\tau}(k{-}q)^\rho
                                   - g^{\lambda\rho}(k'{+}k)^\tau\Bigr)
D^>_{\lambda\nu}(ig (2p{+}k)^\nu t_c)\,
\nonumber \\
&& \hspace*{0.1cm} \times \; 
 D^>_{\tau\sigma}(ig (2p'{-}q)^\sigma t_a)\, 
\frac{-i}{(p{+}k)^2-M^2-i \epsilon}\,  
 D^> (p) (2\pi)^4 \delta^{(4)}(p'{-}p{-}k{-}q)\,
 \frac{d^4p}{(2\pi)^4} \frac{d^4q}{(2\pi)^4} \frac{d^4k}{(2\pi)^4} 
\nonumber \\
&=& \frac{g^4}{2E} f^{cba} t_b t_c t_a 
    \int \frac{d^4q}{(2\pi)^4} \frac{d^4k}{(2\pi)^4} 
    \frac{1}{(p{+}k)^2-M^2-i \epsilon} 
    \, 2\pi\delta(p'_0{-}E{-}k_0{-}q_0) \, G \; ,
\eeqar{M111_1}
where we used Eq.~(\ref{quark_cut}), and performed the integral over $d^4p$. 
As in the previous section, $\vp=\vp'-\vk-\vq$ should be substituted and 
we define $G$ as
\beqar 
G&=& 
(2p'{-}k')^\mu\, (2p{+}k)^\nu\, (2p'{-}q)^\sigma\, D_{\mu\rho}(k')\, 
D^>_{\nu\lambda}(k)\, D^>_{\sigma\tau}(q)\,
\Bigl(g^{\rho\tau}(k'{+}q)^\lambda + g^{\lambda\tau}(k{-}q)^\rho
                                   - g^{\lambda\rho}(k'{+}k)^\tau\Bigr)
\nonumber\\
&=& G_1+G_2-G_3 
\eeqar{G}
with
\beqar
G_1&=&\bigl[(2p'{-}k')_\mu D^{\mu\rho}(k')D^>_{\rho\sigma}(q)\,
(2p'{-}q)^{\sigma}
      \bigr]
      \bigl[(k'{+}q)^\lambda D^>_{\lambda\nu}(k)\,(2p{+}k)^\nu \bigr] , 
\nonumber \\
G_2&=&\bigl[(2p'{-}k')^\mu D_{\mu\rho}(k')\,(k{-}q)^\rho\bigr] 
      \bigl[(2p{+}k)^\nu D^>_{\nu\lambda}(k) D^{> \, \lambda\sigma}(q)\,
            (2p'{-}q)_{\sigma}\bigr] ,
\nonumber \\
G_3&= &\bigl[(2p'{-}k')_\mu D^{\mu\rho}(k') D^>_{\rho\nu}(k)\,(2p{+}k)^{\nu}
      \bigr]
      \bigl[(k{+}k')^\tau D^>_{\tau\sigma}(q)\,(2p'{-}q)^{\sigma}\bigr].
\eeqar{G123}
Recalling that for the radiated gluon only the transverse part of the HTL 
propagator contributes, it is straightforward to show that (under the 
assumptions listed in Appendix~\ref{appa}) the dominant contribution 
to Eq.~(\ref{G}) comes from $G_3$ (i.e. $G_1$ and $G_2$ present small 
corrections which can be neglected). By using 
Eqs.~(\ref{rad_cut}),~(\ref{exchanged_cut}) and 
Eq.~(\ref{Energy_difference}), we obtain
\beqar
 G_3&\approx&  
 \bigl[4p'_\mu D^{\mu\rho}(k') D^>_{\rho\nu}(k)\,p^\nu\bigr]
 \bigl[(k{+}k')^\lambda D^>_{\lambda\sigma}(q)(2p'{-}q)^\sigma\bigr]
\nonumber \\
 &=& \left [-\frac{4 p_z p'_z k_z k'_z}{(k'^2_z{+}\bk'^2)(k_z^2+\bk^2)}\,
     \frac{i \bk\cdot(\bk{+}\bq)}{(k{+}q)^2-m_g^2+i\epsilon}\,
     \frac{2 \pi}{2 \omega} \, \delta ( k_0-\omega) \right]\;\nonumber \\
     &&  \times \left [(1+ f(q_0))
     (4 k_z p'_z)\, \frac{\bq^2}{q_z^2{+}\bq^2} 
     \, 2 {\rm Im} \left(\frac{1}{q^2-\Pi_L(q)} - 
     \frac{1}{q^2-\Pi_T(q)}\right)\right]
\nonumber \\
&\approx& -16 E^2 \, \frac{2 \pi}{2 \omega} \, \delta( k_0-\omega)
\frac{i }{(k+q)^2-m_g^2+i\epsilon}\,
\frac{\bk{\cdot}(\bk{+}\bq)}{x}\, (1+ f(q_0))\, \frac{\bq^2}{q_z^2{+}\bq^2}
\, 2 {\rm Im}\left(\frac{1}{q^2-\Pi_L(q)} - \frac{1}{q^2-\Pi_T(q)}\right).
\eeqar{G3}

Hence, Eq.~(\ref{M111_1}) becomes 
\beqar
M^>_{1,1,1} &\approx& 8E \, g^4\, (i f^{cba} t_b t_c t_a) 
 \int \frac{d^4q}{(2\pi)^4} \frac{d^3k}{(2\pi)^3 2\omega}\,
 \frac{1}{(p{+}k)^2-M^2}\, \frac{1}{(k{+}q)^2-m_g^2} 
 \frac{\bk\cdot(\bk{+}\bq)}{x} 
\nonumber \\
&& \times\, 2\pi \, \delta(p'_0{-}E{-}\omega{-}q_0)\,(1{+}f(q_0)) \, 
   \frac{\bq^2}{q_z^2{+}\bq^2}
   \; 2 {\rm Im} \left(\frac{1}{q^2-\Pi_L(q)} - \frac{1}{q^2-\Pi_T(q)}\right) . 
\eeqar{M111_2}
Noting that $i f^{cba} t_b t_c t_a=\frac{1}{2}[t_a,t_c][t_c,t_a]$, 
the cut amplitude $M_{1,1,1}$ then reads
\beqar
M^>_{1,1,1} &\approx&  4 E \, g^4 \,  [t_a,t_c][t_c,t_a] 
\int \frac{d^3k}{(2\pi)^3 2\omega}\, \frac{d^3q\,dq_0}{(2\pi)^3}\,
     \delta\left(q_0{-}q_z{+}\frac{\bk^2+M^2 x^2 +m_g^2}{2k_z}\right)
     \,(1{+}f(q_0))\,\frac{\bq^2}{q_z^2{+}\bq^2}\,
\\
&\times& 
    \frac{1}{x}\frac{\bk\cdot(\bk{+}\bq)}{((p{+}k)^2-M^2)\,((k{+}q)^2-m_g^2)}\,
          \left(\frac{2\,{\rm Im}\,\Pi_L(q)}
                     {(q^2{-}{\rm Re}\,\Pi_L(q))^2 + ({\rm Im}\,\Pi_L(q))^2}
              - \frac{2\,{\rm Im}\,\Pi_T(q)}
                     {(q^2{-}{\rm Re}\,\Pi_T(q))^2 + ({\rm Im}\,\Pi_T(q))^2}
          \right) .
\nonumber
\eeqar{ImM111}
The $\delta$-function implies that
\beqar
 (k{+}q)^2-m_g^2 \approx -((\bk{+}\bq)^2+M^2 x^2 +m_g^2).
\eeqar{k+q}
With the help of this and Eqs.~(\ref{ppprimek}), (\ref{Energy_difference})
we further obtain
\beqar
M^>_{1,1,1} &\approx& - \, 4 E \, g^4 \, [t_a,t_c][t_c,t_a]
\int \frac{d^3k}{(2\pi)^3 2\omega} \frac{d^3q\,dq_0}{(2\pi)^3}\,
\delta(q_0{-}q_z)\,(1{+}f(q_0))\, 
\frac{\bk\cdot(\bk{+}\bq)}
     {(\bk^2+M^2 x^2 +m_g^2)\,((\bk{+}\bq)^2+M^2 x^2 +m_g^2)}\,
\nonumber \\
&& \hspace*{0.5cm} \; \times 
\frac{\bq^2}{q_z^2{+}\bq^2}
\left(\frac{2\,{\rm Im}\,\Pi_L(q)}
           {(q^2{-}{\rm Re}\,\Pi_L(q))^2 + ({\rm Im}\,\Pi_L(q))^2}
    - \frac{2\,{\rm Im}\,\Pi_T(q)}
           {(q^2{-}{\rm Re}\,\Pi_T(q))^2 + ({\rm Im}\,\Pi_T(q))^2}\right).
\eeqar{M111_4}
Finally, by applying the same procedure as in 
Eqs.~(\ref{f_app})--(\ref{PTLy}), we obtain
\beqar
2\,{\rm Im}\,M_{1,1,1} = -\, 4 E\, g^4 T\, [t_a,t_c][t_c,t_a] 
\int \frac{d^3k}{(2\pi)^3 2\omega}\,\frac{d^2q}{(2\pi)^2}\,
\frac{\bk\cdot(\bk{+}\bq)}
     {(\bk^2+M^2 x^2+m_g^2)\,((\bk{+}\bq)^2+M^2 x^2+m_g^2)}\,
\frac{\mu^2}{\bq^2(\bq^2{+}\mu^2)}\,.\quad
\eeqar{M111_final}
It is straightforward to show that the cut amplitudes of diagrams 
$M_{1,1,2}$, $M_{1,1,3}$, and $M_{1,1,4}$ each lead to the same 
result. The sum of all four diagrams computed in this section thus gives
\beqar
2\,{\rm Im}\,M_{1,1} &\equiv&2\,{\rm Im}\,M_{1,1,1} + 2\,{\rm Im}\,M_{1,1,2}
                           + 2\,{\rm Im}\,M_{1,1,3} + 2\,{\rm Im}\,M_{1,1,4}  
\nonumber \\
& = & 8E\, g^4 T\, [t_a,t_c] [t_c, t_a]
\int \frac{d^3k}{(2\pi)^3 2\omega}\,\frac{d^2q}{(2\pi)^2}\,
\frac{-2\,\bk\cdot(\bk{+}\bq)}
     {(\bk^2+M^2 x^2+m_g^2)\,((\bk{+}\bq)^2+M^2 x^2+m_g^2)}\,
\frac{\mu^2}{\bq^2(\bq^2{+}\mu^2)}\,.
\eeqar{M11_final}

%%%%%%%%%%%%%%%%%%%%%%%%%%%%%%%%%%%%%%%%%%%%%%%%%%%%%%%%%%%%%%%%%%%%%%%%%%%%%
\section{Computation of diagram $\bm{M_{1,2}}$}
\label{appd}
%%%%%%%%%%%%%%%%%%%%%%%%%%%%%%%%%%%%%%%%%%%%%%%%%%%%%%%%%%%%%%%%%%%%%%%%%%%%%

In this Appendix we calculate the diagram shown in Fig.~\ref{DiagM12}
where both ends of the exchanged gluon $q$ are attached to the 
radiated gluon $k$:
%
%%%%%%%%%%%%%%%%%%%%%%%% Fig. 9 %%%%%%%%%%%%%%%%%%%%%%%%%%%%%%%%%%%%%%%%%%%%
\begin{figure}[ht]
\vspace{3.7cm} 
\includegraphics{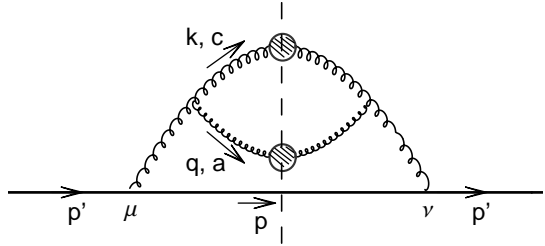}
\caption{Feynman diagram $M_{1,2}$ contributing to the radiative energy loss 
to first order in opacity, labeled in the same way as Fig.~\ref{DiagM10}.}
\label{DiagM12}
\end{figure}
%%%%%%%%%%%%%%%%%%%%%%%%%%%%%%%%%%%%%%%%%%%%%%%%%%%%%%%%%%%%%%%%%%%%%%%%%%%%

%
\beqar
M^>_{1,2} &=& \int (-ig (2p'{-}k')^\mu t_b)\, D_{\mu\rho}(k')\, gf^{bac} 
\Bigl(g^{\rho\tau}(k'{+}q)^\lambda + g^{\lambda\tau}(k{-}q)^\rho
                                   - g^{\lambda\rho}(k'{+}k)^\tau\Bigr)
D^>_{\lambda\alpha}(k) D^>_{\tau\beta}(q) 
\nonumber \\
&& \hspace*{0.1cm} \times\, 
g f^{dac} 
\Bigl(g^{\sigma\beta}(k'{+}q)^\alpha + g^{\alpha\beta}(k{-}q)^\sigma
                                     - g^{\alpha\sigma}(k'{+}k)^\beta\Bigr)
D_{\sigma\nu}^*(k')(ig (2p'{-}q)^\nu t_d)\, D^>(p)\,  
\nonumber \\
&&\hspace*{0.1cm} \times\, 
(2\pi)^4\delta^{(4)}(p'{-}p{-}k{-}q)\,\frac{d^4p}{(2\pi)^4}\,
\frac{d^4q}{(2\pi)^4}\,\frac{d^4k}{(2\pi)^4} 
\nonumber \\
&=& \frac{g^4}{2E}\,f^{bac} t_b f^{dac} t_d 
\int \frac{d^4q}{(2\pi)^4}\,\frac{d^4k}{(2\pi)^4}\,
2\pi\delta(p'_0{-}E{-}k_0{-}q_0)\,H\;,
\eeqar{M12_1}
where we used Eq.~(\ref{quark_cut}), and performed the integral over $d^4p$. 
Again, $\vp=\vp'-\vk-\vq$ should be substituted and we define $H$ as
\beqar 
 H&=& (2p'{-}k')^\mu\,(2p'{-}k')^\nu\ D_{\mu\rho}(k')\,D^>_{\lambda\alpha}(k)\,
      D^>_{\tau\beta}(q)\,D^*_{\sigma\nu}(k')
\nonumber \\
&& \; \; \times 
      \Bigl(g^{\rho\tau}(k'{+}q)^\lambda + g^{\lambda\tau}(k{-}q)^\rho
                                         - g^{\lambda\rho}(k'{+}k)^\tau\Bigr)
      \Bigl(g^{\sigma\beta}(k'{+}q)^\alpha + g^{\alpha\beta}(k{-}q)^\sigma
                                     - g^{\alpha\sigma}(k'{+}k)^\beta\Bigr)\;.
\eeqar{H}
Note that the left and right parts of the $M_{1,2}^>$ diagram are complex 
conjugates (i.e. mirror images) of each other. Therefore, for the 
three-gluon vertices on the left and right side, we go in counter-clockwise 
and clockwise direction, respectively.

As in the previous sections, for the radiated gluon we only consider 
transverse polarization. Under the assumptions described in 
Appendix~\ref{appa}, it is straightforward to show that the dominant 
contribution to Eq.~(\ref{H}) is given by 
\beqar
H&\approx& 
\bigl[4p'_\mu\,D^{\mu\rho}(k')\,D^>_{\rho\sigma}(k)\,(D^{\sigma\nu}(k'))^*\, 
      p'_\nu \bigr]
\bigl[(k'{+}k)^\gamma D^>_{\gamma\eta}(q)\,(k'{+}k)^\eta\bigr]
\nonumber \\
&\approx& \biggl[ \frac{4\,p'^2_z\,\bk'^2}{k'^2_z}\, 
\frac{1}{((k{+}q)^2-m_g^2+i\epsilon)((k{+}q)^2-m_g^2-i\epsilon)} \,
2\pi\frac{\delta(k_0-\omega)}{2\omega} \biggr]
\nonumber \\ 
&& \times \left [4 k'^2_z\, (1+ f(q_0)) \frac{\bq^2}{q_z^2{+}\bq^2} 
     \, 2 {\rm Im} \left(\frac{1}{q^2-\Pi_L(q)} - 
     \frac{1}{q^2-\Pi_T(q)}\right)\right]
\nonumber \\
&\approx& 16 E^2 \frac{(\bk{+}\bq)^2}{((k{+}q)^2-m_g^2)^2}\,
     (1+ f(q_0)) \frac{\bq^2}{q_z^2{+}\bq^2} 
     \, 2 {\rm Im} \left(\frac{1}{q^2-\Pi_L(q)} - 
     \frac{1}{q^2-\Pi_T(q)}\right)\; ,
\eeqar{H1}
where we have used Eqs.~(\ref{rad_cut}),~(\ref{exchanged_cut}) 
and Eq.~(\ref{Energy_difference}).

Therefore, Eq.~(\ref{M12_1}) becomes 
\beqar
M^>_{1,2} &\approx& 8E\,g^4\,f^{bac} t_b f^{dac} t_d 
\int \frac{d^4q}{(2\pi)^4}\,\frac{d^3k}{(2\pi)^3 \, 2 \omega} \,
2\pi\delta(p'_0{-}E{-}\omega{-}q_0)
\nonumber \\ &&\; \; \times
\frac{(\bk{+}\bq)^2}{((k{+}q)^2-m_g^2)^2}\,
(1+ f(q_0)) \frac{\bq^2}{q_z^2{+}\bq^2} 
     \, 2 {\rm Im} \left(\frac{1}{q^2-\Pi_L(q)} - 
     \frac{1}{q^2-\Pi_T(q)}\right) \; .
\eeqar{M12_2}
By using $i f^{bac} t_b = [t_a,t_c]$, Eq.~(\ref{Energy_difference}) and 
Eq.~(\ref{k+q}), we obtain
\beqar
2\,{\rm Im}\,M_{1,2} &\approx& 8E\,g^4\,[t_a,t_c] [t_c,t_a] 
\int \frac{d^4q}{(2\pi)^4}\,\frac{d^3k}{(2\pi)^3 2\omega} 
\frac{(\bk{+}\bq)^2}{((\bk{+}\bq)^2+M^2 x^2+m_g^2)^2}
\frac{\bq^2}{q_z^2{+}\bq^2} \, (1{+}f(q_0)) 
\nonumber \\ &&\; \; \times\, 2\pi\delta(q_0-q_z) 
\left(\frac{2\,{\rm Im}\,\Pi_L(q)}
           {(q^2{-}{\rm Re}\,\Pi_L(q))^2 + ({\rm Im}\,\Pi_L(q))^2}
    - \frac{2\,{\rm Im}\,\Pi_T(q)}
           {(q^2{-}{\rm Re}\,\Pi_T(q))^2 + ({\rm Im}\,\Pi_T(q))^2}\right) \; .
\eeqar{M12_3}
Finally, by applying the same procedure as in 
Eqs.~(\ref{f_app})--(\ref{PTLy}), we obtain
\beqar
2\,{\rm Im}\,M_{1,2} =  8E\,g^4 T\, [t_a,t_c] [t_c t_a] 
\int \frac{d^3k}{(2\pi)^3 2\omega}\,\frac{d^2q}{(2\pi)^2}\,
\frac{(\bk{+}\bq)^2}{((\bk{+}\bq)^2+M^2 x^2+m_g^2)^2} 
\frac{\mu^2}{\bq^2(\bq^2{+}\mu^2)} \,.
\eeqar{M12_final}

%%%%%%%%%%%%%%%%%%% References %%%%%%%%%%%%%%%%%%%%%%%%%%%%%%%%%%%%%%%%%%%%%


\begin{thebibliography}{80}
\bibitem{fn1}
Note that in the widely used static approximation this problem 
is absent: only longitudinally polarized gluons contribute to the static 
potential (see e.g.~\cite{Le_Bellac,Gyulassy_Wang,GLV}), leading to 
individually convergent results for each of the scattering diagrams.
\end{thebibliography}
\end{document}